\documentclass[10pt]{article}
\usepackage[T1]{fontenc}
\usepackage[utf8]{inputenc}
\usepackage{arxiv,microtype}
\usepackage{amsmath,amssymb,amsthm,mathtools,booktabs,graphicx}
\usepackage[round,authoryear]{natbib}
\usepackage{algorithm,algorithmic,authblk,float,flafter}
\usepackage[font=small,labelfont=bf]{caption}
\usepackage[section]{placeins}
\usepackage{xcolor,array,tabularx,longtable}
\usepackage[mathlines]{lineno}
\usepackage[colorlinks=true,allcolors=blue!50!black]{hyperref}
\newcommand{\R}{\mathbb R}
\newcommand{\N}{\mathcal{N}}
\newcommand{\GEV}{\operatorname{GEV}}
\newcommand{\E}{\operatorname{E}}
\newcommand{\Var}{\operatorname{Var}}
\newcommand{\Cov}{\operatorname{Cov}}
\newcommand{\diag}{\operatorname{diag}}

\newcommand{\rh}{\widehat R}
\newcommand{\dd}{\mathrm d}
\newcommand{\ind}{\mathbb I}

\title{Bayesian Inference for Spatial Extremes: Efficient Sampling and Laplace Correction}
\author[1,2]{James Chok\thanks{%
  \href{mailto:james.chok@ed.ac.uk}{james.chok@ed.ac.uk}}%
  \hspace{0.5em}}
\affil[1]{School of Mathematics and Maxwell Institute for
  Mathematical Sciences, University of Edinburgh, United Kingdom}
\affil[2]{School of Mathematics and Statistics, UNSW Sydney,
  Australia}

\date{November 2026}
\renewcommand{\shorttitle}{Bayesian Inference for Spatial Extremes: Efficient Sampling and Laplace Correction}
\hypersetup{
pdftitle={Bayesian Inference for Spatial Extremes: Efficient Sampling and Laplace Correction},
pdfauthor={James Chok}}
\begin{document}
\raggedbottom
\maketitle
% \linenumbers

\begin{abstract}
We exploit latent Gaussian structure of spatial generalized extreme-value models to develop a practical sampler for large spatial scales. Concretely, this sampler obtains five-chain MCMC samples for national rainfall over UK, temperature over France, and snowfall in Canada in minutes to a few hours. Using these samples as MCMC references reveals systematic errors in joint Laplace mode--quadrature (L-MQ) inference in which small, but coherent station errors displace regional return-level distributions. However, we demonstrate that a classical third-order mean correction substantially improves posterior inference for area-averaged return levels, regional contrasts, and exceedance-area curves, often reducing approximation errors by more than an order of magnitude. Importantly, the correction costs less than one second after the initial L-MQ preparation and uses no MCMC input; however, the corrected approximation still understates location uncertainty at some Canadian stations and the spread of the northern French station maximum. Overall, this work provides support for corrected Laplace inference for regional averages and exceedance areas, while establishing that full posterior sampling is a practical option for spatial generalized extreme-value models.
\end{abstract}

\section{Introduction}\label{sec:intro}
Spatial estimation of extremes informs decisions about infrastructure, weather risk, and environmental change. However, station records of rain, temperature, or snow fall are uneven: some sites may have decades of observations, while others contain only a few annual maxima. Bayesian spatial models address this imbalance by borrowing information across neighboring stations, allowing stations with limited observations to benefit from spatial pooling \citep{cooley07, davison2012, sang2009}. Station return levels naturally lead to questions about regional summaries: region-average return values and their changes in climate assessments \citep{Seneviratne2012, Kharin2013, Seneviratne2021}, and the fraction of a region experiencing extreme conditions \citep{Gleason2008, Hansen2012}. In order to answer such questions, one must account for the joint information between return levels across large areas, as opposed to individual station-marginals. This results in a high dimensional posterior, often with funnel-like geometry, and practical tools to fit the model must therefore depend on two questions: how affordable full posterior sampling can be, and whether a faster approximation preserves the spatial quantities being reported.

Impractical computational cost of full posterior sampling has been a large driving factor for approximate inference in this setting, and Bayesian statistics more generally \cite{Lindgren2015}. In particular, \citet{chen2025} considers a toy-example 400-location simulation, with spatial location, scale, and shape fields, for which HMC--NUTS in \texttt{rstan} took 62.5 hours on average across six parallel chains. In a similar fashion, \citet{hazra2025shot} studies a Bangladesh application with 185 grid cells and about 476,000 daily values, reporting 9.3--10.3 hours for 20 parallel chains using a Gibbs sampler with adaptive Metropolis--Hastings and MALA updates. These examples can make posterior approximation methods seem necessary for large station networks. However, as demonstrated by \citet{geirsson2015} and as we show here, designing samplers tailored to the model structure can substantially reduce computational costs, making full posterior inference practical even for large spatial models.

Practical sampling algorithms also change how we assess approximation accuracy. While MCMC estimates posterior quantities by drawing samples, Laplace methods utilize optimization and Gaussian approximations to simplify posterior calculations \citep{rue09}. The joint mode--quadrature (L-MQ) construction of \citet{chen2025} combines Laplace approximations with quadrature over hyperparameters, and their results closely agree with MCMC for station parameters and return levels.

Comparisons between Laplace-based approximations and MCMC for spatial latent Gaussian models have a long history, including log-Gaussian Cox processes \citep{taylor2014, TengNathooJohnson2017}, disease mapping \citep{Carroll2015,VranckxNeyensFaes2019}, ecology \citep{BeguinMartinoRueCumming2012, MonnahanKristensen2018, ClarkDixon2023}, and extended latent Gaussian models \cite{StringerBrownStafford2023, chen2025, hrafnkelsson2021, johannesson2022, opitz2018, vandeskog2022}. However, these studies largely assess parameters and marginal or pointwise summaries, leaving the accuracy of regional estimates less understood. In particular, small but systematic station-level errors may persist and accumulate under spatial averaging, becoming substantial relative to the posterior uncertainty of regional estimates. 

We establish three connected findings:
\begin{itemize}
    \item First, full posterior sampling is computationally practical for large spatial GEV models of UK rainfall, French temperature, and Canadian snowfall, involving up to $7\,907$ unknown parameters. We achieve this through a Laplace-reference split Hamiltonian Monte Carlo (LR-S-HMC) sampler that combines noncentered field updates and centered hyperparameter updates to improve exploration of the hierarchical posterior.
    \item Second, small but systematic L-MQ station-level errors can substantially shift regional return-level distributions, despite close agreement at the station level. A classical third-order mean correction (TMC), giving L-MQ-TMC, removes most of this error without the need of MCMC input, at an additional cost of less than one second after shared Laplace preparation.
    \item Third, correcting the posterior means is sufficient to recover close agreement with MCMC for the studied regional averages, contrasts, and exceedances. However, unsurprisingly, limitations remain for stations exhibiting non-Gaussian posterior geometry and for joint station maxima, whose distributions depend on features beyond the corrected means.
\end{itemize}

Section~\ref{sec:model} defines the applications, model, and posterior targets. Section~\ref{sec:methods} presents the complete computational workflow, and Section~\ref{sec:design} specifies matched comparisons, precision measures, and timing scopes. Section~\ref{sec:results} assesses cost, the localized NUTS comparison, regional inference, and marginal errors. Section~\ref{sec:discussion} discusses method choice for applied work.

\section{Applications, spatial GEV model, and posterior targets}\label{sec:model}
The aim of this paper is to compare inference for a fixed posterior. As such, both the data processing and the model follow established practices rather than being tailored to each application. Concretely, the chosen annual coverage and record-length rules (Section~\ref{sec:data}) match those of \citet{vandeskog2022} and \citet{chen2025}, while the model (Section~\ref{sec:gev-model}) is a standard time-invariant latent Gaussian spatial GEV specification with conditionally independent margins \citep{cooley07, sang2009, geirsson2015, johannesson2022, chen2025}. Specifically, we use an SPDE location and log-scale fields with a common shape, a configuration that SpatialGEV supports directly \citep{SpatialGEV2024}. We \emph{do not} claim that this model best describes each dataset: our conclusions concern computation and approximation for this specification, and choosing a model for each application would require a separate assessment.

\subsection{Rainfall, temperature, and snowfall applications}\label{sec:data}
All three datasets cover 1996--2025 and use the same two-stage filtering rule: accept a calendar year only when at least 70\% of its 365 or 366 days are valid, then retain a station only when at least ten years are accepted. Missing values are not imputed, and accepted years retain their observed maxima rather than an adjustment for missing days.

UK responses are annual maxima of daily rainfall (millimeters) from MIDAS Open \citep{midas2026}; French responses are annual maxima of daily maximum shelter temperature (degrees Celsius) from M\'et\'eo-France \citep{meteofrance2026}, covering metropolitan France and Corsica; and Canadian responses are annual maxima of available monthly snowfall totals in centimeters \citep{ecccdata,eccctech2023}, not maxima of daily snowfall. Table~\ref{tab:data} reports the number of blocks and stations remaining after the two-stage filtering.

\begin{figure}
\centering
\includegraphics[width=\linewidth]
{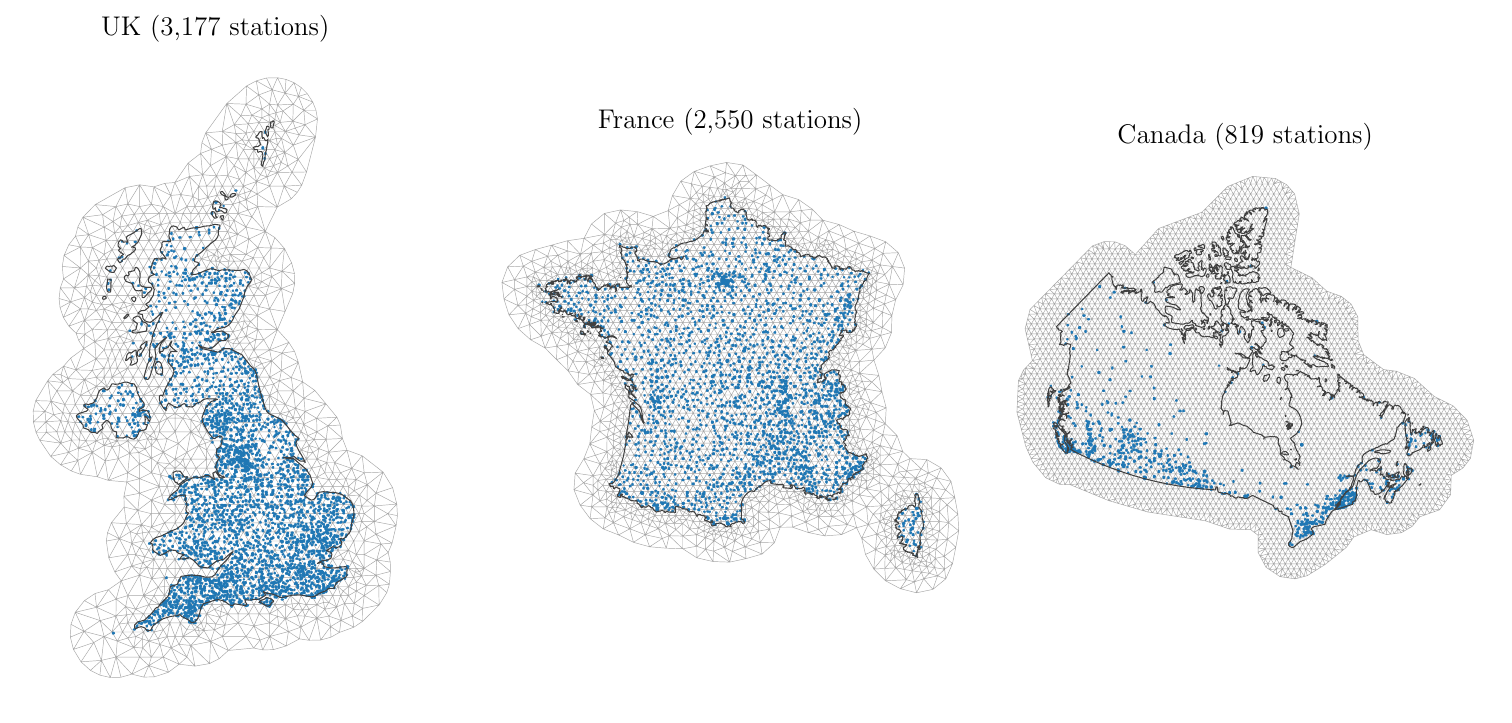}
\caption{UK rainfall, French temperature, and Canadian snowfall meshes. Blue points mark stations; gray lines show triangulations, country outlines, and outer mesh boundaries. The French network includes Corsica.}
\label{fig:meshes}
\end{figure}

Figure~\ref{fig:meshes} shows the locations of the retained stations along with the triangulated meshes used to represent the spatial fields.

\begin{table}
\centering\small
\caption{Cohorts after the quality-control rules of Section~\ref{sec:data}.}
\label{tab:data}
\begin{tabular}{lrrrrrr}
\toprule
Dataset & Stations & Blocks & Years & Median & $m$ & $2m+5$\\
\midrule
UK & 3,177 & 64,190 & 10--30 & 20 & 1,337 & 2,679\\
France & 2,550 & 57,493 & 10--30 & 23 & 2,500 & 5,005\\
Canada & 819 & 13,954 & 10--30 & 15 & 3,951 & 7,907\\
\bottomrule
\end{tabular}
\end{table}

\subsection{Spatial GEV model}\label{sec:gev-model}
Let $Y_{it}$ be the observed annual block maximum at station $\boldsymbol{s}_i$, $i=1,\ldots,n$, in year $t$, and let $\mathcal O$ contain the observed station--year pairs.\footnote{Vectors and matrices use bold symbols, all vectors are columns, and individual entries are unbolded.} Extreme-value theory motivates the GEV distribution for normalized block maxima under appropriate regularity conditions \citep{coles01}. Following \citet{chen2025}, we assume conditional independence given station parameters, with likelihood
\begin{equation}\label{eq:observation}
 L(\boldsymbol{y}\mid\boldsymbol{\mu},\boldsymbol{\eta},\xi)
 =\prod_{(i,t)\in\mathcal O}
 f_{\GEV}(y_{it};\mu_i,e^{\eta_i},\xi).
\end{equation}
Station $i$ has location $\mu_i$ and scale $\sigma_i=e^{\eta_i}$. A single shape parameter $\xi$ applies to every station in the application. Using one shape pools tail information across the short station records and avoids estimating a weakly identified spatial shape field; constant-shape GEV models have an established precedent \citep{sang2009}. Working with log scale $\eta_i$ keeps the scale positive while allowing an unconstrained spatial field. For $a_{it}=(y_{it}-\mu_i)e^{-\eta_i}$ and $b_{it}=1+\xi a_{it}$, the log density is
\begin{equation}\label{eq:gev}
 \log f_{\GEV}(y_{it};\mu_i,e^{\eta_i},\xi)
 =-\eta_i-(1+\xi^{-1})\log b_{it}-b_{it}^{-1/\xi},
 \qquad b_{it}>0,
\end{equation}
with its continuous Gumbel limit at $\xi=0$ and zero density outside the support. Parameter-dependent support requires care in likelihood asymptotics even for independent GEV observations \citep{Smith1985}. Missing blocks contribute no likelihood term; their absence is assumed ignorable for likelihood inference \citep{RUBIN1976}. Parameters are time invariant, and conditional independence applies across stations and years. The model describes spatial variation in marginal extremes, without a separate model for residual dependence between simultaneous events.

To share information across stations, we represent the location and log-scale fields on a triangulated mesh with $m$ vertices and piecewise linear basis functions \citep{lindgren11}. Spatial coordinates are expressed in Kilometers using the British National Grid, Lambert-93, and Statistics Canada Lambert projections for the UK, France, and Canada, respectively.

One issue with the triangulated mesh is that stations generally lie inside the triangles, rather than on the vertices. We recover estimates at the exact station locations through a fixed interpolation matrix $\boldsymbol{A}\in\R^{n\times m}$ which contains the barycentric weights $A_{ij}$. Each row has at most three nonzero entries and sums to one. Write $\boldsymbol{a}_i=(A_{i1},\ldots,A_{im})^\top$ for the transpose of row $i$. The fields and common shape are
\begin{equation}\label{eq:fields}
 \boldsymbol{\mu}=\beta_\mu\boldsymbol{1}_n+\boldsymbol{A}\boldsymbol{v}_\mu,
 \qquad
 \boldsymbol{\eta}=\beta_\eta\boldsymbol{1}_n+\boldsymbol{A}\boldsymbol{v}_\eta,
 \qquad \xi=0.5\tanh\zeta,\quad\zeta\sim\N(0,0.65^2).
\end{equation}
The offsets $\beta_\mu$ and $\beta_\eta$ are fixed prior centers, set by the pooled GEV estimates below, anchoring the fields to countrywide baseline levels. The two fields are independent a priori. We omit an additional station-specific random effect (nugget), matching the GEV-GP field construction of \citet{chen2025}. The shape link restricts $\xi$ to $(-0.5,0.5)$. The upper bound $\xi<0.5$ ensures finite GEV variance at fixed parameters; the lower bound places the individual GEV margins in the regular likelihood regime $\xi>-1/2$ \citep{bucher2017}. The same bounded shape range has precedent in the geophysical prior of \citet{MartinsStedinger2000}, whose beta density differs from our transformed-normal shape prior.

\paragraph{Spatial priors and hyperpriors.}
For $g\in\{\mu,\eta\}$, let $\rho_g>0$ be practical range and $\tau_g>0$ the infinite-domain field SD. The range controls the spatial correlation length, while the field SD controls the size of spatial variation around the fixed prior center. The mass-lumped Mat\'ern SPDE precision in two dimensions uses fixed smoothness $\nu=1$, or SPDE order $\alpha=2$, the standard and most extensively tested case described by \citet{Lindgren2015}:
\begin{equation}\label{eq:spde}
 \begin{split}
 \kappa_g&=\sqrt8/\rho_g,\qquad
 a_g=(\sqrt{4\pi}\kappa_g\tau_g)^{-1},\\
 \boldsymbol{Q}_g&=a_g^2(\kappa_g^2\boldsymbol{M}+\boldsymbol{G})
 \boldsymbol{M}^{-1}(\kappa_g^2\boldsymbol{M}+\boldsymbol{G}),
 \qquad \boldsymbol{v}_g\mid\rho_g,\tau_g
 \sim\N(\boldsymbol{0}_m,\boldsymbol{Q}_g^{-1}).
 \end{split}
\end{equation}
Here $\boldsymbol{M}$ is the diagonal lumped mass matrix and $\boldsymbol{G}$ the stiffness matrix, with natural boundary conditions \citep{lindgren11}. The meshes extend beyond the observation networks to reduce the influence of these boundary conditions on the station-level priors \citep{Lindgren2015,Bakka2018}. The field SD $\tau_g$ is distinct from observation scale $e^{\eta_i}$ and SPDE multiplier $a_g$. Finite-domain nodal variances need not equal $\tau_g^2$.

We use independent penalized-complexity (PC) priors, a standard tail-probability-calibrated construction for Mat\'ern range and field SD \citep{simpson2017,fuglstad2019}, with densities
\begin{equation}\label{eq:pc}
 p(\rho_g)=\lambda_{\rho,g}\rho_g^{-2}e^{-\lambda_{\rho,g}/\rho_g},
 \qquad p(\tau_g)=\lambda_{\tau,g}e^{-\lambda_{\tau,g}\tau_g},
 \qquad \rho_g,\tau_g>0.
\end{equation}
We set $P(\rho_g<\rho_{0,g})=P(\tau_g>\tau_{0,g})=0.5$, giving $\lambda_{\rho,g}=\rho_{0,g}\log2$ and $\lambda_{\tau,g}=\log2/\tau_{0,g}$. The five hyperparameters are
\begin{equation}\label{eq:theta}
 \boldsymbol{\theta}=(\log\rho_\mu,\log\tau_\mu,
 \log\rho_\eta,\log\tau_\eta,\zeta)^\top.
\end{equation}
Together with the $2m$ field coefficients, these give $2m+5$ free parameters (Table~\ref{tab:data}). Their density includes the range and SD log-coordinate Jacobians; the normal prior is specified directly on $\zeta$. UK and French calibration thresholds are $(\rho_{0,\mu},\rho_{0,\eta})=(300,300)$\,km and $(\tau_{0,\mu},\tau_{0,\eta})=(15,0.5)$. The Canadian mesh uses $(600,600)$\,km and $(40,1.0)$. Location field SD uses the response unit, while log-scale field SD is dimensionless. For each country, we fit one GEV with constant location, scale, and shape by maximum likelihood to all selected observed station--year maxima, giving each observed block equal weight. The pooled location estimate and log of the pooled scale estimate give $(\beta_\mu,\beta_\eta)=(32.916,2.385)$ for the UK, $(33.619,1.379)$ for France, and $(40.052,3.263)$ for Canada. The pooled shape is a nuisance parameter of this fit; the spatial model retains the shape prior in \eqref{eq:fields}. These data-derived prior centers introduce an empirical-Bayes element: all comparisons condition on their plug-in values without propagating offset-estimation uncertainty.

\subsection{Posterior quantities of interest}\label{sec:targets}
A local return level summarizes the upper tail of the modeled block distribution. For return period $T>1$, define
\begin{equation}\label{eq:return}
 z_T(\boldsymbol{s})=\mu(\boldsymbol{s})
 +c_T(\xi)e^{\eta(\boldsymbol{s})},\qquad
 c_T(\xi)=\frac{\{-\log(1-T^{-1})\}^{-\xi}-1}{\xi},
\end{equation}
using the continuous limit at zero shape. This is the $(1-1/T)$ quantile of the modeled annual block distribution. Away from stations, the same mesh basis defines $\mu(\boldsymbol{s})$ and $\eta(\boldsymbol{s})$; we interpolate log scale before exponentiating. Our main comparisons use $T=100$.

For a mapped land region $R$ with physical area $|R|$, define
\begin{equation}\label{eq:regional}
 \overline z_R=\frac1{|R|}\int_R z_{100}(\boldsymbol{s})\,\dd A,
 \qquad
 A_R(X)=\frac1{|R|}\int_R
 \ind\{z_{100}(\boldsymbol{s})>X\}\,\dd A.
\end{equation}
Region-averaged return values are standard climate-assessment summaries \citep{Seneviratne2012,Kharin2013,Li2021}; Canadian assessments also report regional changes in 20-year return values \citep{ECCC2016}. Regional frequency analysis provides a related UK rainfall precedent through GEV growth curves \citep{FowlerKilsby2003}, although a growth curve is not itself an area average of local return levels. Area fractions beyond thresholds underlie operational extremes indices \citep{Gleason2008} and studies of land affected by heat extremes \citep{Hansen2012,CoumouRobinson2013}. Those studies use thresholds relative to local climatology. Here $X$ is a return-level threshold: $A_R(X)$ traces the spatial distribution of $z_{100}$ within $R$.

Regions follow external reporting partitions: the four UK nations used in Met Office climate series \citep{MetOfficeRegionalSeries}, the 13 metropolitan French r\'egions, and the six Canadian regions in Chapter~1, Figure~1.1 of \citet{BushLemmen2019}. The Canadian partition is British Columbia, Prairies, Ontario, Quebec, Atlantic, and Canada's North (Yukon, Northwest Territories, and Nunavut). We examine the four UK nations, Hauts-de-France and Provence-Alpes-C\^ote d'Azur, and Atlantic and Prairies in detail. Appendix~\ref{app:full-partitions} reports scalar summaries for every region in each full partition. For each posterior draw, we also calculate the same-draw contrast $\overline z_{R_1}-\overline z_{R_2}$, retaining all cross-region covariances. All six UK pairs and one pair each in France and Canada are assessed.

For each regional average and paired contrast, we report the posterior CDF, median, and central 95\% credible interval. For the exceedance-area fraction $A_R(X)$, we report its posterior mean and pointwise 95\% credible interval at each threshold $X$. Neither target is a return level of a spatially aggregated observation or the footprint probability of a simultaneous event.

\section{Posterior computation}\label{sec:methods}
Write $\boldsymbol v=(\boldsymbol v_\mu^\top,\boldsymbol v_\eta^\top)^\top$ for the two mesh fields. For $g\in\{\mu,\eta\}$, let $\boldsymbol h_g=(\log\rho_g,\log\tau_g)^\top$ and collect the four spatial log hyperparameters in $\boldsymbol h=(\boldsymbol h_\mu^\top,\boldsymbol h_\eta^\top)^\top$. Conditional on the observed maxima, the fixed mesh, and the fitted offsets, the joint posterior is
\begin{equation}\label{eq:joint-posterior}
\begin{aligned}
 \pi(\boldsymbol v,\boldsymbol h,\zeta\mid\boldsymbol y)
 &\propto L(\boldsymbol y\mid\boldsymbol\mu,\boldsymbol\eta,\xi)
           \,p_\zeta(\zeta) \\
 &\quad{}\times
 \prod_{g\in\{\mu,\eta\}}
 \left[
 \phi_m(\boldsymbol v_g;\boldsymbol 0_m,\boldsymbol Q_g^{-1})
 \,p(\rho_g)p(\tau_g)\rho_g\tau_g
 \right].
\end{aligned}
\end{equation}
Here $\phi_m$ denotes the $m$-variate Gaussian density, $\boldsymbol Q_g$ is evaluated at $\boldsymbol h_g$, and $p_\zeta$ is the $\N(0,0.65^2)$ density. The factors $\rho_g\tau_g$ come from expressing the priors in log coordinates. The station parameters and $\xi$ follow from \eqref{eq:fields}.

We use two features of \eqref{eq:joint-posterior} for computation. At fixed $\boldsymbol h$, the Gaussian field priors admit a noncentered representation. At fixed $(\boldsymbol v,\zeta)$, the observation likelihood is constant in $\boldsymbol h$. LR-S-HMC uses these properties to update the posterior in two blocks. L-MQ constructs a joint Gaussian approximation, whose field means we subsequently correct.

\subsection{Practical MCMC}\label{sec:mcmc}
The sampler must move thousands of spatial coefficients together while also exploring their dependence on four prior hyperparameters and a common shape. We use a finite-adaptive Laplace-reference split-HMC sampler, abbreviated LR-S-HMC. Its ingredients are Gaussian-reference split Hamiltonian dynamics \citep{beskos2011,shahbaba2014,ottobre2016} with noncentered field moves and centered hyperparameter moves \citep{papaspiliopoulos2007}. This combination exploits the latent Gaussian structure to make full posterior sampling practical for the fitted models. The Gaussian references determine proposal geometry, whereas every acceptance test uses the specified supported posterior.

Related structure-exploiting methods include the LGM split sampler, which partitions latent variables according to their role in the likelihood \citep{geirsson2020}, and transport-map HMC using approximate conditional geometry \citep{osmundsen2021}. These algorithms differ from the split Hamiltonian block updates used here.

A full fixed-mesh modal decomposition provides square roots $\boldsymbol{B}_g(\boldsymbol{h}_g)$ satisfying $\boldsymbol{B}_g\boldsymbol{B}_g^\top=\boldsymbol{Q}_g^{-1}$, and we set $\boldsymbol{v}_g=\boldsymbol{B}_g\boldsymbol{z}_g$, with $\boldsymbol{z}_g\sim\N(\boldsymbol{0}_m,\boldsymbol{I}_m)$, for $g\in\{\mu,\eta\}$. All $m$ modes are retained: this is a coordinate change for the fitted mesh rather than a low-rank approximation. The first block updates the innovations and shape coordinate, $\boldsymbol{x}=(\boldsymbol{z}_\mu^\top, \boldsymbol{z}_\eta^\top,\zeta)^\top$, conditional on $\boldsymbol{h}$. Its target is
\begin{equation}\label{eq:mcmc_field_target}
 \pi(\boldsymbol{x}\mid\boldsymbol{h},\boldsymbol{y})
 \propto L(\boldsymbol{y}\mid\boldsymbol{\mu},\boldsymbol{\eta},\xi)
 \exp\left\{-\tfrac12\sum_g\|\boldsymbol{z}_g\|^2
 -\frac{\zeta^2}{2(0.65)^2}\right\}.
\end{equation}
The likelihood uses the mesh-to-station projection. Noncentering removes the direct prior-scale dependence of the innovations, but the likelihood still couples them to the hyperparameters.

After this move, the sampler changes representation and updates $\boldsymbol{h}$ while holding the physical fields $(\boldsymbol{v}_\mu,\boldsymbol{v}_\eta)$ fixed. The likelihood is then constant, and the centered conditional target is
\begin{equation}\label{eq:mcmc_hyper_target}
 \pi(\boldsymbol{h}\mid\boldsymbol{v},\zeta,\boldsymbol{y})
 \propto\prod_g |\boldsymbol{Q}_g(\boldsymbol{h}_g)|^{1/2}
 \exp\{-\tfrac12\boldsymbol{v}_g^\top\boldsymbol{Q}_g
 \boldsymbol{v}_g\}\,p(\rho_g)p(\tau_g)\rho_g\tau_g.
\end{equation}
The last factors are the log-coordinate Jacobians. Updating the four hyperparameters in this representation avoids evaluating the observation likelihood during their trajectories. Transforming back to innovations under the accepted hyperparameters preserves the physical fields. An iteration consists of these two block moves.

Both blocks use the same split-dynamics construction with separate references and acceptance tests. For a block target $\pi_b(\boldsymbol{q})$ and frozen Gaussian reference $\N(\boldsymbol{c},(\boldsymbol{L}\boldsymbol{L}^\top)^{-1})$, whiten with $\boldsymbol{w}=\boldsymbol{L}^\top (\boldsymbol{q}-\boldsymbol{c})$. In these coordinates the negative log target is $\|\boldsymbol{w}\|^2/2+\mathcal R(\boldsymbol{w})$, up to a constant. A fresh standard normal momentum $\boldsymbol{p}$ is followed by symmetric residual kicks and exact rotations of the Gaussian position--momentum pair. After $K$ steps through total angle $\Theta$, the endpoint $\boldsymbol{q}'$ is accepted with probability
\begin{equation}\label{eq:mcmc_accept}
 \alpha_b=1\wedge\exp\left\{
 \log\pi_b(\boldsymbol{q}')-\log\pi_b(\boldsymbol{q})
 -\tfrac12(\|\boldsymbol{p}'\|^2-\|\boldsymbol{p}\|^2)\right\}.
\end{equation}
Adjacent half kicks are merged, giving $K+1$ force evaluations per block. For the field block only, a deterministic continuation supplies forces near or outside the GEV support. All prescribed steps are completed, and unsupported or nonfinite endpoints are rejected using the supported target. The continuation therefore changes the proposal rather than the inferential model. This differs from replacing the likelihood by a likelihood with parameter-independent support, such as the blended GEV \citep{castrocamilo2022}. Appendix~\ref{app:sampler} gives the transformations, split map, and numerical safeguards.

\begin{algorithm}[htbp]
\caption{Finite-adaptive LR-S-HMC workflow and one retained iteration}
\label{alg:mcmc}
\begin{algorithmic}[1]
\REQUIRE Fixed data, mesh, offsets, and priors; five chain states and separate random-number streams.
\STATE Construct the full modal basis and conditional Gaussian references for the field/shape and centered hyperparameter blocks. Initialize supported chain states by separate perturbations of the reference center.
\STATE During discarded warmup, adapt the two shared block angles and refresh references at the prescribed window boundaries. Fit the field/shape reference at pooled median log hyperparameters; do not change live chain states during fitting.
\STATE Freeze both references and angles, then run the fixed settling phase. Discard all warmup and settling states.
\FOR{each retained iteration, separately within each chain}
\STATE At current $\boldsymbol{h}$, draw fresh field momentum and propose $\boldsymbol{x}=(\boldsymbol{z}_\mu^\top, \boldsymbol{z}_\eta^\top,\zeta)^\top$ with split dynamics. Accept or reject using \eqref{eq:mcmc_field_target} and \eqref{eq:mcmc_accept}.
\STATE Form $\boldsymbol{v}_g=\boldsymbol{B}_g(\boldsymbol{h}_g) \boldsymbol{z}_g$, $g=\mu,\eta$, from the resulting field state.
\STATE At fixed $(\boldsymbol{v}_\mu,\boldsymbol{v}_\eta,\zeta)$, draw fresh hyperparameter momentum and propose $\boldsymbol{h}'$ with split dynamics. Apply a separate acceptance test using \eqref{eq:mcmc_hyper_target} and \eqref{eq:mcmc_accept}.
\STATE Under the accepted hyperparameters, set $\boldsymbol{z}_g=\boldsymbol{B}_g(\boldsymbol{h}_g)^{-1}\boldsymbol{v}_g$. Retain the state, including repeated block values after rejection.
\ENDFOR
\end{algorithmic}
\end{algorithm}

% References and angles adapt only during discarded warmup
% \citep{chok2025twosystem}; Table~\ref{tab:sampler_tuning} gives the
References and angles adapt only during discarded warmup; Table~\ref{tab:sampler_tuning} gives the schedule and frozen settings. Retained transitions and Canadian continuations use frozen references and angles.

The computational benefit of the SPDE construction does not make this sampler wholly sparse. Its full modal eigendecomposition requires dense setup; modal transforms, field/shape reference Hessians, Cholesky factors, and whitening factors are also dense. The modal basis has $O(m^2)$ storage and $O(m^3)$ setup, and dense transforms recur during sampling. The centered four-dimensional move is cheap by comparison. We batch five chains in JAX float64 \citep{jax2018github} on a ten-core Apple M1 Pro CPU. These implementation choices explain the practical costs for the UK and France and the larger investment for Canada; the timing comparison concerns this implementation and the reported runs.

\subsection{Joint L-MQ approximation}
For fixed hyperparameters $\boldsymbol{\theta}$, let $\widehat{\boldsymbol{v}}(\boldsymbol{\theta})$ be the conditional field mode and $\boldsymbol{H}(\boldsymbol{\theta})$ the Hessian of the conditional negative log posterior at that mode. The Hessian describes the local curvature: directions with greater curvature have less posterior variation under a Gaussian approximation. Its inverse gives the corresponding conditional covariance. Laplace integration uses this local Gaussian to approximate the hyperparameter posterior by
\begin{equation}\label{eq:laplace}
 p_{\rm LA}(\boldsymbol{\theta}\mid\boldsymbol{y})
 \propto p\{\boldsymbol{y},
 \widehat{\boldsymbol{v}}(\boldsymbol{\theta}),\boldsymbol{\theta}\}
 |\boldsymbol{H}(\boldsymbol{\theta})|^{-1/2}.
\end{equation}
At its mode $\boldsymbol{\theta}^\star$, define $\widehat{\boldsymbol{v}}=\widehat{\boldsymbol{v}} (\boldsymbol{\theta}^\star)$, $\boldsymbol{H}= \boldsymbol{H}(\boldsymbol{\theta}^\star)$, $\boldsymbol{V}=\{-\nabla^2_{\boldsymbol{\theta}}\log p_{\rm LA} (\boldsymbol{\theta}^\star\mid\boldsymbol{y})\}^{-1}$, and the mode sensitivity $\boldsymbol{J}=\partial\widehat{\boldsymbol{v}}/ \partial\boldsymbol{\theta}^\top|_{\boldsymbol{\theta}^\star}$. The sensitivity describes how the conditional field mode moves as the hyperparameters change. It therefore allows hyperparameter uncertainty to propagate into the fields without finding a new mode for every draw. The joint mode--quadrature reconstruction \citep{chen2025} is
\begin{equation}\label{eq:mq}
 \boldsymbol{\theta}\sim\N(\boldsymbol{\theta}^\star,\boldsymbol{V}),
 \qquad
 \boldsymbol{v}_{\text{L-MQ}}=\widehat{\boldsymbol{v}}
 +\boldsymbol{J}(\boldsymbol{\theta}-\boldsymbol{\theta}^\star)
 +\boldsymbol{\epsilon},\qquad
 \boldsymbol{\epsilon}\sim\N(\boldsymbol{0}_{2m},\boldsymbol{H}^{-1}),
\end{equation}
with independent hyperparameter and residual innovations. Thus $\E_{\text{L-MQ}}\boldsymbol{v}=\widehat{\boldsymbol{v}}$ and $\Var_{\text{L-MQ}}(\boldsymbol{v})=\boldsymbol{H}^{-1} +\boldsymbol{J}\boldsymbol{V}\boldsymbol{J}^\top$. The first covariance term represents variation at fixed hyperparameters; the second represents variation from the changing conditional mode. This variance propagation is related to classical parametric empirical-Bayes approximations \citep{kass1989}. The reconstruction retains the field mean at the original mode, which can differ from the posterior mean when the conditional posterior is skewed. Nonlinear transformations, including $e^{\eta_i}$, shape $\xi$, and return levels, need not be Gaussian under this reconstruction.

On synthetic datasets, we compare the JAX and TMB Gaussian reconstructions using identical observations, finite-element matrices, offsets, and priors (Appendix~\ref{app:validation}). We assess posterior accuracy against MCMC.

\subsection{Third-order mean correction, L-MQ-TMC}
A Gaussian approximation is symmetric about its mode, so its curvature alone cannot describe a displacement between the mode and the mean. Third derivatives describe changes in curvature and supply the leading local term accounting for this asymmetry.

Write $\Phi(\boldsymbol{v},\boldsymbol{\theta}) =-\log p(\boldsymbol{v}\mid\boldsymbol{y},\boldsymbol{\theta})$, up to terms independent of $\boldsymbol{v}$. At $\boldsymbol{\theta}^\star$, let $\boldsymbol{\Sigma}=\boldsymbol{H}^{-1}$. The classical local Laplace moment expansion \citep{lindley1980,tierney1989,rue09} gives the leading mean shift below. Higher-order Laplace corrections have also reduced parameter bias in spatio-temporal count models \citep{ClarkDixon2023}. Specifically,
\begin{equation}\label{eq:t3_shift}
 \delta_j=-\frac12\sum_{a,b,c=1}^{2m}
 \Sigma_{ja}T_{abc}\Sigma_{bc},\qquad
 T_{abc}=\partial_a\partial_b\partial_c\Phi
 (\widehat{\boldsymbol{v}},\boldsymbol{\theta}^\star).
\end{equation}
The negative sign follows from expanding the negative log posterior; Appendix~\ref{app:t3-derivation} derives the contraction from Gaussian fourth moments. We call this third-order mean correction because it uses third derivatives in the leading local moment term; the name does not imply third-order accuracy for the entire posterior distribution.

The contraction has the same third-derivative structure as the location and skewness corrections in INLA's simplified Laplace approximation (SLA) \citep{rue09}. It also agrees to first local order with the fixed-covariance Gaussian variational Bayes (VB) mean condition \citep{vanniekerk2024}. For $q=\N(\widehat{\boldsymbol{v}}+\boldsymbol{\delta},\boldsymbol{\Sigma})$, expanding the formal stationarity condition $\E_q[\nabla\Phi(\boldsymbol{v})]=\boldsymbol{0}$ around the conditional mode gives
\[
 \boldsymbol{H}\boldsymbol{\delta}
 +\tfrac12(\boldsymbol{T}:\boldsymbol{\Sigma})\simeq\boldsymbol{0},
 \qquad (\boldsymbol{T}:\boldsymbol{\Sigma})_a
 =\sum_{b,c}T_{abc}\Sigma_{bc},
\]
and hence \eqref{eq:t3_shift}. This is a local correspondence of the mean terms. A full Gaussian assigns mass outside nonzero-shape GEV support, so its reverse Kullback--Leibler divergence from the exact conditional posterior, $\operatorname{KL}(q\Vert p)$, is infinite. Variational optimization therefore requires an additional treatment of that support.

Other corrections address different parts of Gaussian-based inference: copula correction within INLA \citep{ferkingstad2015}, variational corrections to joint latent means \citep{vanniekerk2024}, extensions to marginal variance and skewness with a Gaussian copula \citep{dutta2026}, and importance-sampling corrections to integrated Laplace marginalization \citep{lai2026}. We evaluate the mean corrections below; the other approaches are not tested here.

For linear field functionals at fixed $\boldsymbol{\theta}^\star$, $\boldsymbol{\delta}$ also coincides with the $\varepsilon$-method bias correction of \citet{ThorsonKristensen2016}, available in TMB via \texttt{sdreport(..., bias.correct = TRUE)}. Differentiating the Laplace approximation to $\log\int\exp\{-\Phi(\boldsymbol{v}) +\varepsilon\boldsymbol{e}_j^\top\boldsymbol{v}\}\,\dd\boldsymbol{v}$ at $\varepsilon=0$ gives $\widehat v_j+\delta_j$. This is an identity between approximations, not an exact posterior-mean formula. TMB stores corrected expectations of requested quantities separately, without recentering its Gaussian approximation; SpatialGEV leaves this option off. Here, after the station covariance blocks are available, one final sparse solve corrects every field coordinate, and translation defines a joint approximation for intervals and spatial functionals. Nonlinear return-level means in \eqref{eq:analytic_mean} are not $\varepsilon$-method expectations, which include additional derivatives of the requested quantity.

The station likelihood factorization makes the contraction practical. At fixed shape, each station contribution depends on just its location and log scale, so its third derivatives can be evaluated locally. The Gaussian prior contributes to the Hessian but has zero third derivative. We can therefore evaluate the correction without constructing a field-sized third-order tensor. Let $\boldsymbol{P}_i=\diag(\boldsymbol{a}_i^\top,\boldsymbol{a}_i^\top)$ project the two fields to station $i$, and define
\begin{equation}\label{eq:t3_station_blocks}
 \boldsymbol{C}_i=\boldsymbol{P}_i\boldsymbol{\Sigma}\boldsymbol{P}_i^\top,
 \qquad
 T_{i,abc}=-\partial_a\partial_b\partial_c
 \ell_i(\widehat\mu_i,\widehat\eta_i;\xi^\star),
 \quad a,b,c\in\{\mu,\eta\}.
\end{equation}
Here $\xi^\star=0.5\tanh\zeta^\star$. Each station log likelihood sums the contributions of its observed blocks. We combine its third derivatives with the conditional station covariance by defining $t_{i,a}=\sum_{b,c}T_{i,abc}C_{i,bc}$ and assemble
\begin{equation}\label{eq:t3_sparse_solve}
 \boldsymbol{b}=\sum_i\boldsymbol{P}_i^\top\boldsymbol{t}_i,
 \qquad
 \boldsymbol{H}\boldsymbol{\delta}=-\tfrac12\boldsymbol{b}.
\end{equation}
Each interpolation row has at most three nonzero entries. Local $2\times2\times2$ derivatives and station covariance blocks therefore suffice. Projection maps these station contributions back to the mesh, and the final solve propagates them through the coupled fields using the original sparse field factor. Thus, local derivative calculations still produce a spatially coupled correction. Mixed location--log-scale derivatives correct both field means jointly.

\paragraph{Variational comparison, L-MQ-VBC.}\label{sec:vbc}
As a numerical comparison, L-MQ-VBC minimizes the Gaussian expected negative log likelihood plus the Gaussian-prior quadratic over all $2m$ location and log-scale mean shifts, holding the conditional covariance and hyperparameter law fixed \citep{vanniekerk2024}. A finite $C^2$ likelihood continuation makes the Gaussian mean objective well defined despite the support mismatch; it is a surrogate objective, not a finite reverse KL to the supported posterior. Appendix~\ref{app:vbc-implementation} gives its implementation. After shared L-MQ preparation this fitting protocol adds 7.18, 13.65, and 6.38 seconds for the UK, France, and Canada, excluding continuation sensitivity checks and posterior assessment. The continuation matters empirically: for the Provence-Alpes-C\^ote d'Azur area average, VBC's CDF discrepancy is 0.036 at width 0.02, changing to 0.090 and 0.020 at widths 0.01 and 0.04 with the remaining settings fixed. TMC uses derivatives of the supported likelihood at the mode and requires no continuation or optimization.

\paragraph{Joint reconstruction and scope.}
We denote the approximation with third-order mean correction by L-MQ-TMC:
\begin{equation}\label{eq:t3_reconstruction}
 \boldsymbol{v}_{\text{L-MQ-TMC}}=\widehat{\boldsymbol{v}}
 +\boldsymbol{\delta}
 +\boldsymbol{J}(\boldsymbol{\theta}-\boldsymbol{\theta}^\star)
 +\boldsymbol{\epsilon}.
\end{equation}
The shift is calculated once at modal hyperparameters. Neither $\boldsymbol{H}$, $\boldsymbol{J}$, $\boldsymbol{V}$, nor the shape hyperparameter law is refitted, and the shift's dependence on hyperparameters is not further linearized. All latent Gaussian covariances are retained. Return-level variances can nevertheless change: adding a constant to log scale multiplies scale by a constant, changing the variation in its contribution to the return level.

Write $z_{100,i}=z_{100}(\boldsymbol{s}_i)$ and let $\boldsymbol{d}_g=\boldsymbol{A}\boldsymbol{\delta}_g$ and $\gamma_i=\E_{\text{L-MQ}}[c_{100}(\xi)e^{\eta_i}]$. The paired reconstructions satisfy
\begin{equation}\label{eq:analytic_mean}
 \begin{split}
 z_{100,i}^{\text{L-MQ-TMC}}-z_{100,i}^{\text{L-MQ}}
 &=d_{\mu,i}+(e^{d_{\eta,i}}-1)c_{100}(\xi)e^{\eta_i},\\
 \E_{\text{L-MQ-TMC}}z_{100,i}-\E_{\text{L-MQ}}z_{100,i}
 &=d_{\mu,i}+(e^{d_{\eta,i}}-1)\gamma_i.
 \end{split}
\end{equation}
Here $\eta_i$ on the right is the L-MQ draw. The first identity holds draw by draw, at stations or at any spatial point using the same mesh basis, and after physical-area integration; the second takes expectations under these finite-moment Gaussian reconstructions. Gaussian exponential tilting and one-dimensional integration over shape evaluate the station means without sampling noise. Regional CDFs, quantiles, contrasts, and exceedance fractions use paired joint draws and physical-area integration.

\section{Comparison design and numerical reliability}\label{sec:design}
We use the three national applications to assess posterior approximation error and a separate rainfall model around Edinburgh to compare two samplers. Within each national application, LR-S-HMC, L-MQ, L-MQ-TMC, and L-MQ-VBC use the same observed station--year maxima in the same station order, the same mesh and station interpolation weights, the same fixed location and log-scale offsets, and the same field and shape priors. The underlying observation model is the GEV likelihood in \eqref{eq:observation}. VBC uses a smooth continuation of its log likelihood when fitting the mean shift. For each country, we apply the TMC and VBC shifts to the same 20,000 L-MQ draws and compare all three approximations with the LR-S-HMC draws from that country's mesh.

\subsection{Localized Edinburgh sampler comparison}\label{sec:matched}
To compare LR-S-HMC with MAP-preconditioned Stan NUTS on the same rainfall posterior, we extract a mesh around Edinburgh from the national UK mesh. The local mesh contains 100 vertices and 162 original triangles. Its 277 stations contribute 5,493 annual maxima under the UK coverage and record-length rules for 1996--2025. Reassembling the FEM matrices imposes natural boundary conditions on the extracted domain; exterior observations are excluded. The local model uses the UK priors and offsets, so this defines a new finite-domain posterior, rather than a subset of the full UK posterior. Both samplers target this same posterior, but their updates use different parameterizations. NUTS jointly updates the fully noncentered vector; LR-S-HMC combines noncentered field/shape moves with centered hyperparameter moves at fixed physical fields. Four chains retain 5,000 iterations each. LR-S-HMC uses the UK settings with 3,000 discarded iterations; MAP-preconditioned Stan NUTS \citep{hoffman2014nuts,carpenter2017stan} uses 5,000 warmup iterations.\footnote{We ran NUTS with 3,000 warmup iterations; however, the total run time was about 5 minutes longer than using 5,000 warmup iterations. This is likely because the adapted preconditioning matrix poorly represents the posterior geometry, forcing NUTS to explore deeper trees during sampling.} Appendix~\ref{app:nuts} specifies the preconditioning, tuning, and validation protocol.

Both Edinburgh runs report wall-clock time on the same ten-core CPU. NUTS runs its four chains as parallel single-threaded processes, so its sampling time is that of the slowest chain; LR-S-HMC advances its four chains as one batched JAX computation. Neither run was restricted to a fixed number of cores, so the timing ratios compare wall time on a common machine rather than equal CPU cost.

\subsection{MCMC reliability}
Rank-normalized split $\widehat R$, bulk ESS, and tail ESS are screened at $\widehat R<1.01$ and both ESS values at least 400 \citep{vehtari2021}. The inventory includes innovations, transformed hyperparameters, and physical mesh fields, so the number of checked coordinates exceeds the number of free parameters. These are finite-run screens, not guarantees of convergence. We initialize the five national chains near a reference. During warmup, we pool information across the chains to adapt the block angles and refresh the field reference. During retained sampling, each chain uses a separate random-number stream with the same frozen references and angles. No separately adapted, overdispersed full-country replication is available.

UK and France begin with 5,000 retained iterations per chain, while Canada begins with 10,000. Each adds 1,000 if any screen fails. UK and France have a 10,000-iteration cap; the Canadian cap is 20,000.

Monte Carlo standard errors (MCSEs) quantify retained-run precision. For station field means, we project physical mesh draws to stations and use the larger of the 250-draw nonoverlapping batch SE and the SE of the five chain means. Batch means account for serial dependence when estimating Monte Carlo variance \citep{flegal2010}. Appendix~\ref{app:reference-precision} reports the precision scales. We additionally screen rank-based diagnostics for the areal averages, contrasts, and all exceedance-grid coordinates, without extending the fitted chains.

For the localized Edinburgh sampler comparison, we require rank-normalized split $\widehat R<1.01$ and bulk and tail ESS of at least 400 for each sampler. We also report the number of divergent NUTS transitions. For each of the five hyperparameters, we compare the posterior median and the endpoints of the central 95\% credible interval on its natural scale. These and the station/network means, SDs, and endpoints are compared relative to combined autocorrelation-aware MCSEs; batch estimates assess sensitivity of these errors. For all summary comparisons, the precise estimators and multiplicity adjustment are described in Appendix~\ref{app:nuts}.

\subsection{Accuracy measures}\label{sec:accuracy}
For scalar target $g$, reference posterior $\pi$, and approximation $q$, we compare posterior means and spreads when these moments exist. Standardizing the mean error by the reference SD expresses its size relative to posterior uncertainty, while the SD ratio compares the two spreads. Define
\begin{equation}\label{eq:metrics}
 D_q(g)=\frac{\E_qg-\E_\pi g}{\operatorname{SD}_\pi(g)},\qquad
 S_q(g)=\frac{\operatorname{SD}_q(g)}{\operatorname{SD}_\pi(g)}.
\end{equation}
Agreement corresponds to mean error zero and SD ratio one. Network mean absolute error (MAE) averages $|D_q|$ across stations.

Let $Q_p^{(q)}(g)$ be the $p$th quantile of $g$ under $q$, with $Q_p^{(\pi)}(g)$ defined analogously. For the central 95\% credible endpoints, define standardized errors
\begin{equation}\label{eq:endpoint-errors}
 E_{q,p}(g)=\frac{Q_p^{(q)}(g)-Q_p^{(\pi)}(g)}
 {\operatorname{SD}_\pi(g)},
 \qquad p\in\{0.025,0.975\}.
\end{equation}
We summarize their absolute values over stations. To assess how these intervals align with posterior mass, we measure how much reference posterior mass lies below and above the approximate interval. Define
\begin{equation}\label{eq:content}
 L_q(g)=P_\pi\{g<Q_{.025}^{(q)}(g)\},\quad
 U_q(g)=P_\pi\{g>Q_{.975}^{(q)}(g)\}.
\end{equation}
For continuous distributions the intended tail contents are 0.025. Atoms at zero or one in an exceedance fraction can give smaller strict tail contents even for an exact interval. We flag either content exceeding 0.05 and report both sides, since eliminating upper-tail flags alone does not establish accurate intervals.

For whole distributions, we compare cumulative probabilities. The maximum CDF discrepancy measures the largest difference between the approximate and reference CDFs:
\begin{equation}\label{eq:distances}
 K_q(g)=\sup_x|F_q(x)-F_\pi(x)|.
\end{equation}
These are posterior approximation diagnostics for matched discretized models. The CDF discrepancy is measured on the probability scale. The two tail probabilities describe reference posterior mass for the observed data, rather than frequentist coverage across repeated datasets or predictive calibration.

For the Gaussian approximations, we calculate station means from Gaussian moments. For the MCMC reference, we estimate them from the retained draws. CDF discrepancies, quantiles, and regional functionals use all retained joint draws. For areal averages and contrasts, we report $K$, medians, credible endpoints, and the reference tail probabilities $L$ and $U$. For $g=A_R(X)$, we evaluate \eqref{eq:metrics} and \eqref{eq:content} at each saved threshold. Figures~\ref{fig:area-difference}--\ref{fig:area-profiles} specify the common plotting coordinate and pointwise bands.

Empirical CDF differences include Monte Carlo variation and do not constitute independent-draw hypothesis tests. For the 23 physical-area averages, eight contrasts, and station-maximum diagnostics, our common-law benchmark compares 20,000 IID rows from the pooled reference with an independently circular-block-resampled reference in each of 200 replicates, using within-chain blocks of length 250. The resulting ranges are central 95\% replicate ranges and pointwise conditional yardsticks, not simultaneous intervals or tests of posterior truth (Appendices~\ref{app:areal-precision} and~\ref{app:precision}). The maximum comparisons also use paired approximation draws and within-chain block resampling to quantify variation in their empirical $K$ values.

\subsection{Timing conventions}\label{sec:timing}
National MCMC times include sampler setup, warmup, and all retained continuations. Approximation times distinguish L-MQ fitting, shared preparation, incremental correction, and posterior assessment. TMC begins from the shared L-MQ mode, factor, and station covariance and includes derivative evaluation and the sparse solve. Polygon integration is posterior assessment and is excluded from fitting times. These component times do not establish a matched-precision end-to-end speedup.

\section{Results}\label{sec:results}
For each country, Appendix~\ref{app:parameter-maps} maps the MCMC means of location and scale, reports the mean shape, and compares all three parameters with L-MQ and L-MQ-TMC. We first assess the cost and numerical reliability of full posterior sampling in the three national applications and compare LR-S-HMC with Stan NUTS on the Edinburgh rainfall model. We then use the national MCMC draws to ask how station-level approximation errors affect regional return-level distributions, how much of that error a correction to the field means removes, and which features of the posterior remain difficult to recover.

\subsection{Affordable full posterior sampling}\label{sec:affordable}
We first assess if full posterior sampling provides computationally practical reference for each national application at the mesh resolution used for inference. We compare the five-chain LR-S-HMC runs using elapsed sampler time and the diagnostics for both the sampled coordinates and the physical mesh fields.

All three national runs pass the sampler-coordinate and physical-field screens, the UK and France after 5,000 retained iterations per chain and Canada after 13,000 (Appendix~\ref{app:canada-continuations}). Complete sampler calls take 13.2 minutes for the UK, 41.9 minutes for France, and 3.70 hours for Canada (Table~\ref{tab:cost}).

\begin{table}[htbp]
\centering\small
\caption{Five-chain MCMC totals and approximation fitting components. Draws are retained per chain; MCMC minutes sum complete sampler-call timers including warmup and all continuations. Diagnostic extrema cover sampler coordinates and natural-scale checks. L-MQ and incremental TMC seconds exclude shared draw generation and posterior assessment.}
\label{tab:cost}
\begin{tabular}{lrrrrrrr}
\toprule
Dataset & Draws & MCMC min & $\max\rh$ & Min bulk & Min tail & MQ s & TMC s\\
\midrule
UK & 5,000 & 13.23 & 1.0013 & 1881 & 3325 & 15.60 & 0.509\\
France & 5,000 & 41.92 & 1.0011 & 5626 & 8971 & 57.29 & 0.536\\
Canada & 13,000 & 222.15 & 1.0082 & 583 & 791 & 59.68 & 0.502\\
\bottomrule
\end{tabular}
\end{table}

All three national fits reached the prescribed sampling criteria on the reported CPU, including the 7,907-parameter Canadian model. This made it feasible to assess the approximations against MCMC at the mesh resolution used in each application. L-MQ fitting took under one minute per country, and TMC added about half a second once the L-MQ factor and station covariance blocks were available.

\subsection{Comparison with MAP-preconditioned NUTS}
\label{sec:nuts-comparison}
On the localized Edinburgh rainfall model (Section~\ref{sec:matched}), we compare LR-S-HMC and MAP-preconditioned Stan NUTS in two respects. We examine agreement in the reported posterior summaries at the available Monte Carlo precision and measure the effective sample size each implementation delivers per second of elapsed time.

For the reported NUTS run, we set the target acceptance to 0.995. Runs at target acceptances of 0.8, 0.9, and 0.95 give retained divergence rates of 5.6\%, 1.7\%, and 1.4\% (Table~\ref{tab:nuts-pilots}), and the reported cost includes the additional integration.

Posterior summaries are broadly consistent at the measured Monte Carlo precision. None of the 3,347 comparisons across five hyperparameters, station quantities, and station-network summaries survives the specified Holm correction. Some pointwise differences remain: the largest is the location mean at Kinross, at $3.62$ combined MCSEs, with adjusted $p=0.981$. The station means and SDs closely follow the identity lines in Figure~\ref{fig:nuts-posterior-agreement}; full hyperparameter comparisons and MCSE sensitivity checks are in Appendix~\ref{app:nuts}. These results support numerical consistency without establishing posterior equivalence.

\begin{table}[htbp]
\centering\small
\setlength{\tabcolsep}{4pt}
\caption{ Edinburgh comparison in wall-clock seconds; NUTS sampling time is that of the slowest chain. Diagnostic extrema cover 205 common sampler coordinates and 838 reported natural-scale or derived quantities. Medians and ESS rates use only the 205 common coordinates; ESS may exceed the retained count under negative autocorrelation. Divergences are not applicable to LR-S-HMC. }
\label{tab:nuts-cost}
\begin{tabular*}{\linewidth}
{@{\extracolsep{\fill}}lrrrrrr@{}}
\toprule
Method & \shortstack{Total\\(s)} & \shortstack{Sampling\\(s)} & $\max\widehat R$ & \shortstack{Min.\\bulk ESS} & \shortstack{Min.\\tail ESS} & Div.\\
\midrule
LR-S-HMC & 75.29 & 19.41 & 1.0016 & 2705 & 3106 & N/A\\
NUTS & 1775.82 & 456.27 & 1.0016 & 3059 & 2897 & 39\\
\bottomrule
\end{tabular*}
\par\smallskip
\textit{Common-coordinate efficiency}\par
\smallskip
\begin{tabular*}{\linewidth}
{@{\extracolsep{\fill}}lrrrrrr@{}}
\toprule
Method & \shortstack{Median\\bulk ESS} & \shortstack{Median\\tail ESS} & \shortstack{Bulk ESS\\/total s} & \shortstack{Tail ESS\\/total s} & \shortstack{Bulk ESS\\/sampling s} & \shortstack{Tail ESS\\/sampling s}\\
\midrule
LR-S-HMC & 13495 & 14336 & 179.24 & 190.41 & 695.34 & 738.68\\
NUTS & 27478 & 15346 & 15.47 & 8.64 & 60.22 & 33.63\\
\bottomrule
\end{tabular*}
\end{table}

Both methods pass the prescribed rank-$\widehat R$ and bulk/tail ESS screens (Table~\ref{tab:nuts-cost}). Median bulk/tail ESS per whole-job second is 11.58/22.03 times larger for LR-S-HMC. NUTS has larger median ESS at the same retained length, so equal draw counts do not imply equal precision.

The local data contain about 20 annual blocks per station. When station likelihoods are informative, full noncentering can induce inverse-funnel dependence between innovations and hyperparameters \citep{papaspiliopoulos2007}. This is a plausible contributor to the NUTS divergences, rather than an identified cause. The hyperparameter bulk ESS per whole-job second is 37.8 times larger for LR-S-HMC, compared with 11.0--11.4 for station quantities (Table~\ref{tab:nuts-efficiency}). This pattern is consistent with a benefit from the centered hyperparameter block. The experiment does not isolate that benefit from Gaussian-reference splitting or other implementation choices.

On this Edinburgh fit, LR-S-HMC delivered higher median bulk and tail ESS per second of total wall time. The posterior summaries were consistent at the precision assessed by the multiplicity-adjusted comparisons. Both samplers passed the rank and ESS screens, although NUTS recorded 39 divergent retained transitions. No NUTS transition reached the tree-depth limit.

\FloatBarrier
\subsection{Coherent errors and regional inference}\label{sec:regional-results}
Small errors in station return levels can survive spatial averaging when they have the same sign across a region. We examine this possibility for three summaries of the local 100-year return-level field. We compare its average over a region, the difference between two regional averages computed on the same posterior draw, and the fraction of a region above a specified return-level threshold.

\begin{figure}[htbp]
\centering
\includegraphics[width=\linewidth]{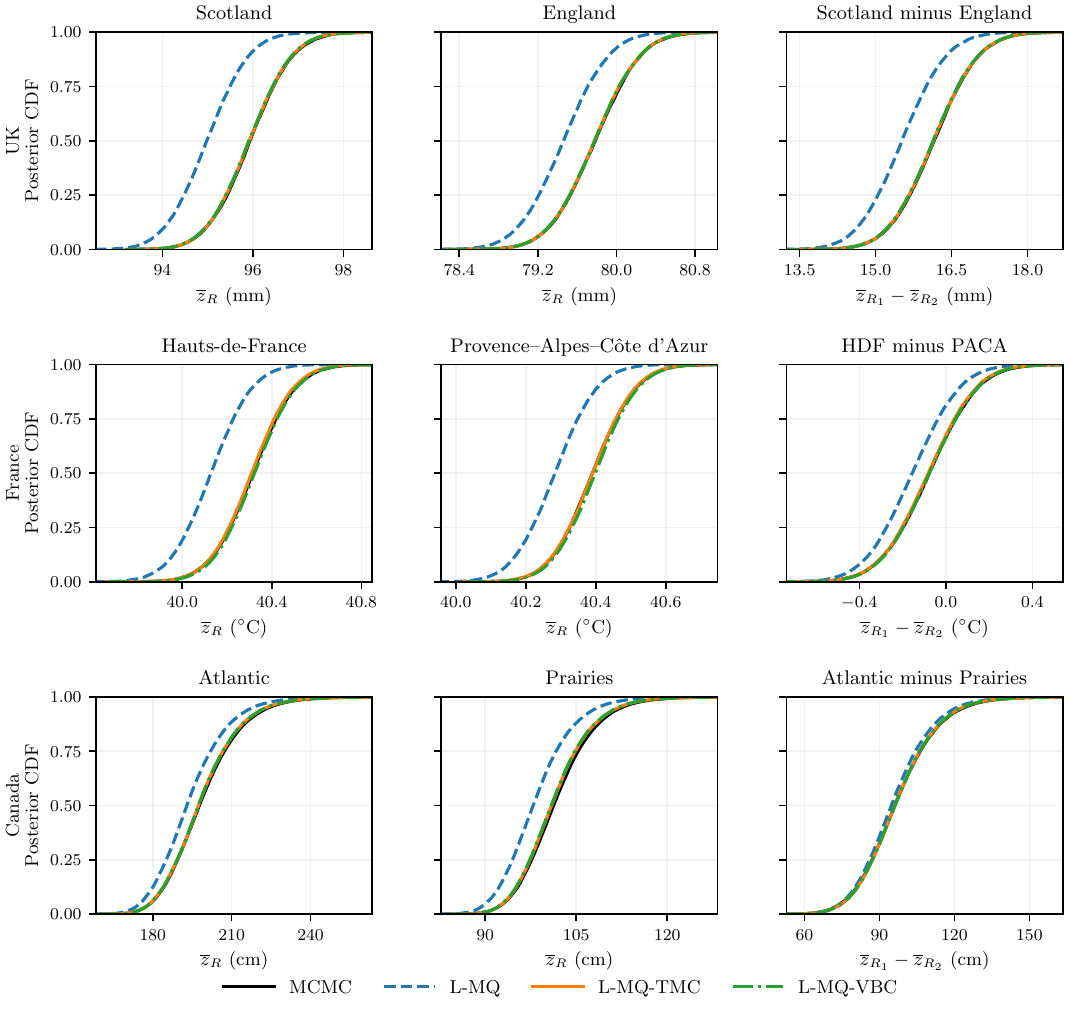}
\caption{Posterior CDFs of area-averaged local 100-year return levels and paired contrasts. Rows show the UK (Scotland, England), France (Hauts-de-France, HDF; Provence-Alpes-C\^ote d'Azur, PACA), and Canada (Atlantic, Prairies); the third column is the first region minus the second. Axes use response units, without posterior-moment standardization. Horizontal limits cover all methods' 0.1st--99.9th percentiles.}
\label{fig:regional}
\end{figure}

L-MQ's station return-level mean errors are modest, 0.11--0.21 posterior SDs on average (Table~\ref{tab:station-accuracy}), yet the largest area-average CDF discrepancy in each country lies between 0.28 and 0.63 (Table~\ref{tab:regions}), peaking at 0.633 in Nouvelle-Aquitaine. Across all 23 regional averages, the largest CDF discrepancy is 0.031 for L-MQ-TMC, in Canada's North, and 0.036 for L-MQ-VBC, in Provence-Alpes-C\^ote d'Azur. For TMC, ten regional averages exceed their respective finite-pool Monte Carlo benchmarks. All four UK averages and all six UK contrasts fall within theirs (Appendix~\ref{app:areal-precision}). Figure~\ref{fig:regional} displays two region-average distributions and their paired contrast per country.

Coherent errors can partly cancel in a contrast, so a small contrast error alone does not establish accurate regional levels (Table~\ref{tab:regional-accuracy}). MCMC gives posterior probability 0.336 of a positive Hauts-de-France minus Provence-Alpes-C\^ote d'Azur contrast.

Quantile-based interval checks make the regional consequences tangible. For Nouvelle-Aquitaine, 44.6\% of the MCMC reference lies above L-MQ's upper 95\% interval endpoint; TMC reduces this to 2.8\%. Neither corrected method has a tail-content flag for any area average or contrast (Table~\ref{tab:areal-tails}).

\begin{table}[htbp]
\centering\small
\caption{Regional accuracy over the eight detailed regions. The first two columns maximize CDF discrepancy over region averages and all within-country paired contrasts. The remaining columns summarize exceedance-area profiles only where $0.02\le p\le0.98$; $N$ is the number of correlated region--threshold coordinates. Counts are not independent tests. All $L>0.05$ counts are zero.}
\label{tab:regional-accuracy}
\begin{tabular}{llrrrrrr}
\toprule
Dataset & Method & $\max K(\overline z_R)$ & $\max K(\mathrm{contrast})$
& $\max|D(A)|$ & $S(A)$ range & $U>.05$ & $N$\\
\midrule
UK & L-MQ & .473 & .346 & 1.049 & .852--1.174 & 127 & 149\\
UK & L-MQ-TMC & .013 & .011 & .036 & .971--1.009 & 0 & 149\\
UK & L-MQ-VBC & .016 & .015 & .041 & .970--1.009 & 0 & 149\\
France & L-MQ & .466 & .176 & 1.148 & .791--1.151 & 109 & 109\\
France & L-MQ-TMC & .019 & .015 & .141 & .935--1.021 & 0 & 109\\
France & L-MQ-VBC & .036 & .007 & .116 & .943--1.022 & 0 & 109\\
Canada & L-MQ & .242 & .044 & .915 & .883--1.070 & 47 & 81\\
Canada & L-MQ-TMC & .028 & .006 & .083 & .966--1.003 & 0 & 81\\
Canada & L-MQ-VBC & .032 & .006 & .092 & .965--1.003 & 0 & 81\\
\bottomrule
\end{tabular}
\end{table}

\begin{figure}[p]
\centering
\includegraphics[width=\linewidth]{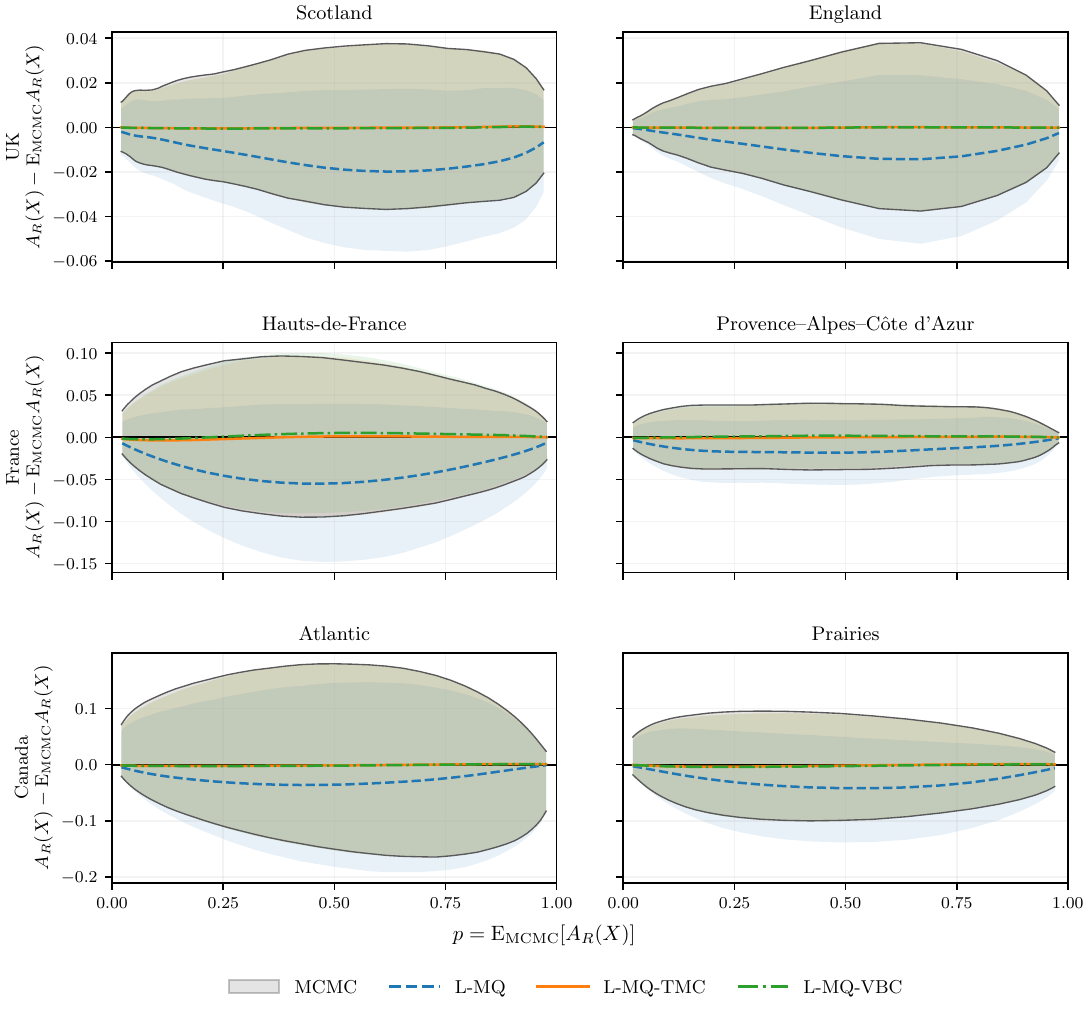}
\caption{Exceedance-area differences for the six regions in Figure~\ref{fig:regional}. Every method's mean and pointwise 95\% credible endpoints are recentered by the MCMC mean at the same physical threshold $X$. The gray reference interval is recentered in the same way. The common horizontal coordinate is $p=\E_\pi[A_R(X)]$; only saved thresholds with $0.02\le p\le0.98$ are shown. A recentered equal-tail band need not contain zero when its distribution is skewed. Bands are pointwise posterior intervals, not simultaneous bands or MCSEs.}
\label{fig:area-difference}
\end{figure}

\begin{figure}[p]
\centering
\includegraphics[width=\linewidth]{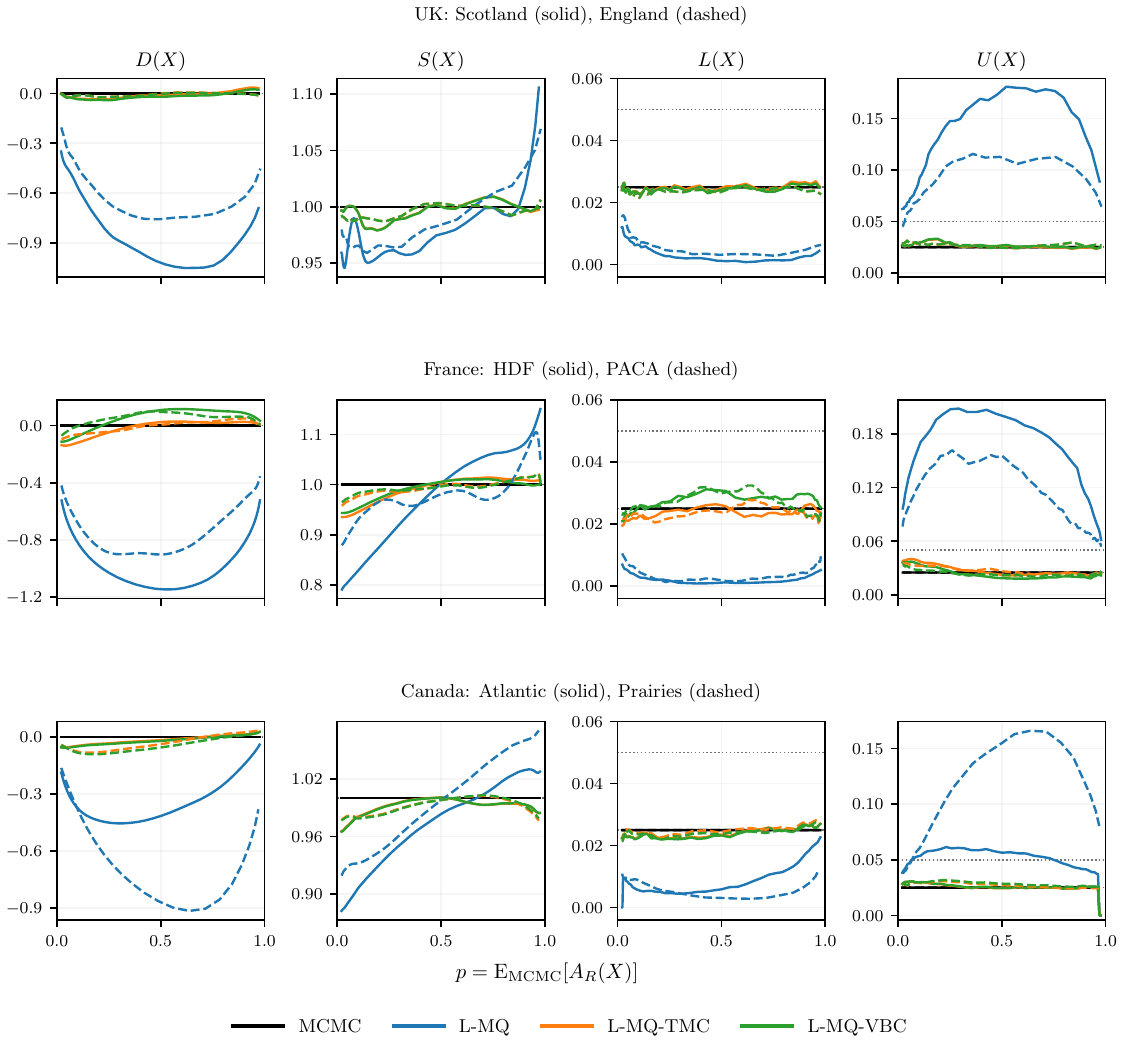}
\caption{Exceedance-area discrepancy profiles at the same thresholds as Figure~\ref{fig:area-difference}. $D$ and $S$ follow \eqref{eq:metrics}; $L$ and $U$ are strict reference tail contents beyond the approximate 95\% endpoints. Reference lines mark zero mean error, unit SD ratio, nominal continuous tail content 0.025, and the 0.05 tail flag. Curves compare methods at a common physical $X$ within each region; the displayed $p$ axis does not equate thresholds between regions. HDF and PACA abbreviate Hauts-de-France and Provence-Alpes-C\^ote d'Azur. Black curves show the empirical MCMC tail contents. Zero reference variance leaves $D$ and $S$ undefined and is shown as a gap.}
\label{fig:area-profiles}
\end{figure}

Exceedance areas expose the same errors in a bounded reported quantity. Figures~\ref{fig:area-difference} and \ref{fig:area-profiles} show centered intervals and the $D,S,L,U$ profiles on the common $p$ axis. Across the central range, the largest $|D|$ falls from 1.15 for L-MQ to 0.141 for L-MQ-TMC. L-MQ produces upper-tail flags at most central coordinates. Both L-MQ-TMC and L-MQ-VBC clear all these flags (Table~\ref{tab:regional-accuracy}).

Agreement at individual stations can therefore conceal a substantial displacement of a regional return-level distribution. Subtracting two regional averages can conceal part of the displacement again when both move in the same direction. Across these applications, both mean corrections brought the regional intervals and exceedance-area profiles closer to MCMC.

\FloatBarrier
\subsection{Explaining and correcting the centering error}
\label{sec:mean-correction}
A local return level depends on location and on the exponential of log scale, so an error in either field can displace a regional average. To identify which field accounts for the improvement under TMC, we compare the calculated station mean shifts with the shifts needed to match MCMC and decompose the change in each regional average into location and scale contributions. Across the three countries, TMC reduces the network mean absolute standardized error in station log-scale means from 0.126--0.254 to 0.006--0.009 (Table~\ref{tab:station-accuracy}), near the reference MCSE scale (Appendix~\ref{app:reference-precision}).

\begin{table}[htbp]
\centering\small
\caption{Network mean absolute standardized errors in station field means, return-level means, and 95\% return-level endpoints. $L,U$ count stations whose corresponding reference tail content exceeds 0.05. $S_\mu<0.9$ counts underdispersed location marginals.}
\label{tab:station-accuracy}
\begin{tabular}{llrrrrrrrr}
\toprule
Dataset & Method & $\mu$ & $\eta$ & $z_{100}$ & $q_{.025}$ & $q_{.975}$ & $L$ & $U$ & $S_\mu<.9$\\
\midrule
UK & L-MQ & 0.012 & 0.126 & 0.111 & 0.114 & 0.156 & 0 & 6 & 0\\
UK & L-MQ-TMC & 0.006 & 0.006 & 0.006 & 0.023 & 0.038 & 0 & 0 & 0\\
France & L-MQ & 0.039 & 0.254 & 0.194 & 0.215 & 0.320 & 0 & 664 & 0\\
France & L-MQ-TMC & 0.007 & 0.009 & 0.009 & 0.050 & 0.105 & 0 & 0 & 0\\
Canada & L-MQ & 0.033 & 0.241 & 0.209 & 0.186 & 0.358 & 0 & 282 & 52\\
Canada & L-MQ-TMC & 0.008 & 0.008 & 0.012 & 0.031 & 0.106 & 0 & 0 & 52\\
\bottomrule
\end{tabular}
\end{table}

By \eqref{eq:analytic_mean}, we can separate the change in each area-averaged draw into its location and scale contributions. Table~\ref{tab:attribution} gives the decomposition for the six regions shown in Figure~\ref{fig:regional}.

\begin{table}[htbp]
\centering\small
\caption{Area-weighted attribution of the paired L-MQ-TMC minus L-MQ change for the six regions in Figure~\ref{fig:regional}. Location contributions are exact; scale and total entries average the 20,000 paired approximate draws. Values are in response units (UK: mm; France: degrees Celsius; Canada: cm). These are changes between finite-moment approximations, not MCMC posterior means.}
\label{tab:attribution}
\begin{tabular}{llrrr}
\toprule
Dataset & Region & Location & Scale & Total\\
\midrule
UK & Scotland & $-0.0402$ & $+0.9798$ & $+0.9396$\\
UK & England & $-0.0066$ & $+0.3236$ & $+0.3170$\\
France & Hauts-de-France & $-0.0216$ & $+0.2030$ & $+0.1814$\\
France & Provence-Alpes-C\^ote d'Azur & $-0.0128$ & $+0.1204$ & $+0.1076$\\
Canada & Atlantic & $-0.2742$ & $+4.7430$ & $+4.4688$\\
Canada & Prairies & $-0.0897$ & $+3.1449$ & $+3.0552$\\
\bottomrule
\end{tabular}
\end{table}

\begin{figure}[htbp]
\centering
\includegraphics[width=\linewidth]
{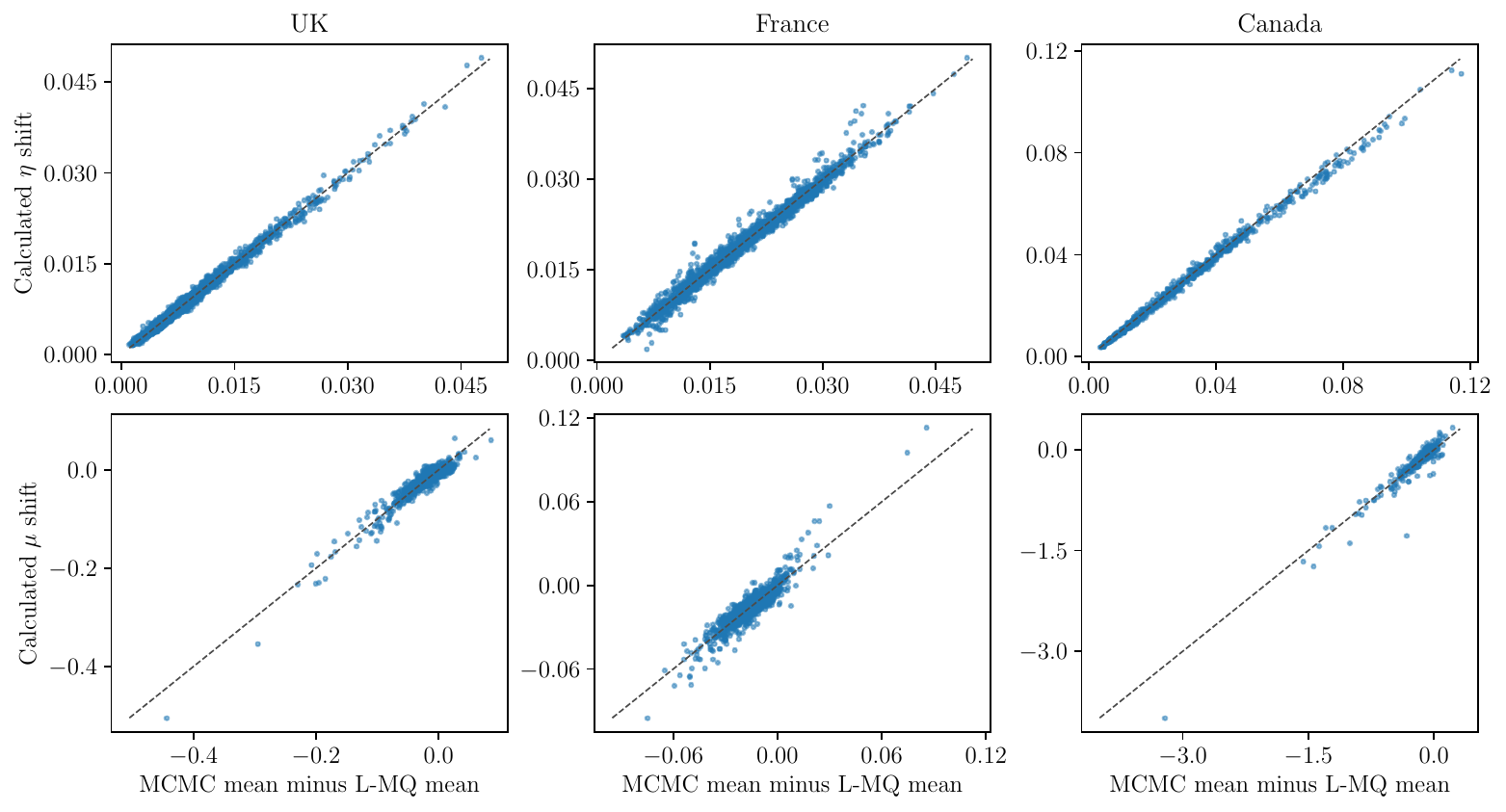}
\caption{Calculated third-order mean shifts versus shifts required to match the MCMC means. Rows show log scale and location; columns show the three applications. Dashed lines are identities. Location shifts use mm, degrees Celsius, and cm; log-scale shifts are dimensionless. Each point is a retained station. For $g\in\{\mu,\eta\}$, the calculated shift is $d_{g,i}=(\boldsymbol A\boldsymbol\delta_g)_i$, the required shift is $\delta^{\mathrm{ref}}_{g,i}=\E_{\mathrm{MCMC}}[g(\boldsymbol s_i)] -\E_{\mathrm{MQ}}[g(\boldsymbol s_i)]$, and the corrected residual is $d_{g,i}-\delta^{\mathrm{ref}}_{g,i} =\E_{\mathrm{TMC}}[g(\boldsymbol s_i)] -\E_{\mathrm{MCMC}}[g(\boldsymbol s_i)]$.}
\label{fig:correction-comparison}
\end{figure}

The calculated shifts closely track those needed to match MCMC station means (Figure~\ref{fig:correction-comparison}).

\FloatBarrier
In all six regions shown in Table~\ref{tab:attribution}, the log-scale shift raises the area-averaged return level, and the smaller location shift partly offsets that increase. This explains why correcting the field means can move an entire regional distribution appreciably. The improvement also extends to station return-level intervals. After TMC, no station has more than 5\% of the MCMC reference draws beyond either endpoint of its approximate 95\% interval (Table~\ref{tab:station-accuracy}).

\subsection{Remaining error in joint posterior summaries}
\label{sec:remaining-error}
\begin{figure}[htbp]
\centering
\includegraphics[width=\linewidth]
{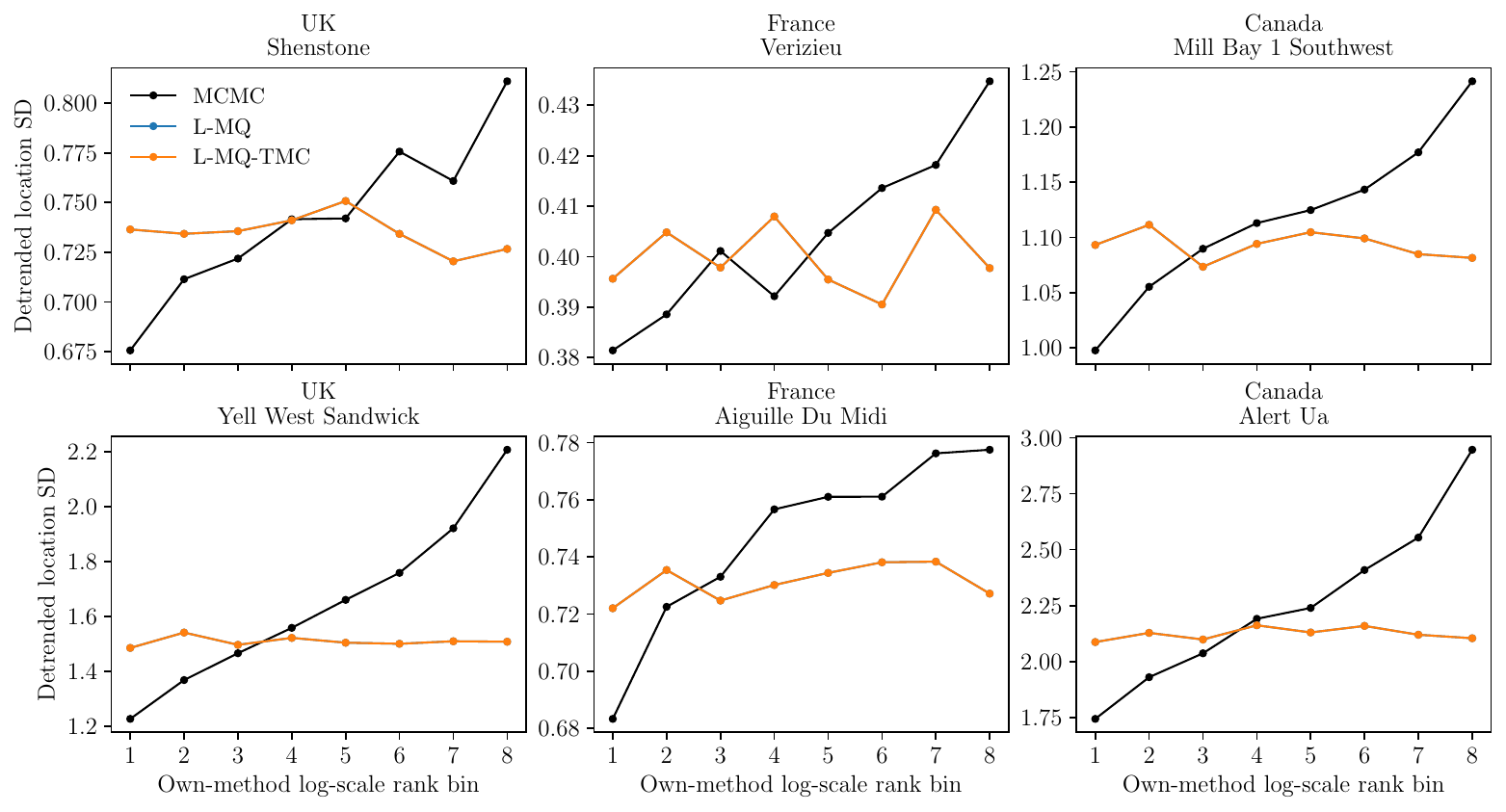}
\caption{Detrended location SD within each method's eight equal-count log-scale rank bins at metadata-selected sites. Black is MCMC, blue L-MQ, and orange L-MQ-TMC. Orange overlays blue because the correction is a deterministic translation. Location uses mm, degrees Celsius, and cm for the UK, France, and Canada. The sites are the shortest accepted history and northernmost station (highest-altitude station for France), selected from metadata before posterior comparison. The binned spread is descriptive, not the conditional variance at a fixed log scale; the sites do not give a network-wide frequency estimate.}
\label{fig:geometry}
\end{figure}

TMC translates the latent fields while preserving their Gaussian covariance. We examine the limits of that correction through two summaries that depend on posterior spread. The first measures variation in station location across log-scale bins (Figure~\ref{fig:geometry}). The second takes the maximum local 100-year return level across a set of northern stations on each posterior draw (Appendix~\ref{app:station-maxima}).

In Canada, 52 of 819 location marginals have SD ratios below 0.9 under both L-MQ and L-MQ-TMC (Table~\ref{tab:station-accuracy}). At the UK station Yell West Sandwick (17 years), MCMC's upper-to-lower-bin SD ratio is 1.80, versus 1.02 for the Gaussian reconstruction. At the Canadian station Alert UA (26 years), the ratios are 1.69 and 1.01. The record-length restriction therefore does not remove every relevant departure from joint Gaussian geometry.

\FloatBarrier
The northern French stations illustrate the distinction between recovering station means and recovering the distribution of a maximum. TMC gives a largest station mean of $43.288\,^\circ\mathrm C$, close to the MCMC value of $43.290\,^\circ\mathrm C$, yet the SD of the draw-by-draw maximum is only 0.871 times its MCMC value (Table~\ref{tab:maximum-uplift}). Together with the Canadian location results, this shows why uncertainty in a reported maximum or at an individual station warrants a direct check even when the corrected means agree closely.

\subsection{Summary of the experimental findings}
Full posterior sampling was feasible on all three national meshes and provided a reference for assessing the quantities reported in each application. Against that reference, the main error in uncorrected L-MQ was a systematic displacement of the log-scale field that persisted under spatial averaging. Correcting the field means substantially improved regional averages, paired contrasts, and exceedance-area profiles. TMC achieved these improvements at an incremental cost of about half a second per fit, and VBC gave broadly similar regional results. The remaining discrepancies were most evident in the spread of some Canadian station locations and of the northern French station maximum. Across these experiments, the accuracy of a regional average was therefore a poor guide to the accuracy of a maximum.

\section{Discussion}\label{sec:discussion}
The national fits and localized Edinburgh comparison (Sections~\ref{sec:affordable} and \ref{sec:nuts-comparison}) make full posterior sampling a practical option for these applications. Latent Gaussian structure supplies computational opportunities through split dynamics and blocked parameterizations. The Edinburgh efficiency pattern is consistent with a benefit from centered hyperparameter moves, showing why the complete computational strategy matters for cost as well as posterior agreement.

Accessible MCMC also makes the consequences of approximation an empirical question. Coherent negative log-scale errors in L-MQ shift regional distributions even when typical station mean errors are modest. Their common direction persists under spatial averaging, whose uncertainty is often more concentrated than station uncertainty. Exceedance areas show the consequence for a bounded reporting quantity. This pattern connects earlier close agreement for station parameters \citep{chen2025} with the less uniform agreement for random-effect quantities in other spatial applications \citep{Carroll2015}.

L-MQ-TMC repairs most of this error. The exact translation identity identifies the log-scale shift as the dominant areal return-level contribution, partly offset by location. Quantile-based $L,U$ checks show close interval content after correction, although Canada's North retains a CDF discrepancy above its conditional Monte Carlo benchmark.

The full-rank, fixed-covariance L-MQ-VBC comparison gives broadly similar regional CDFs to TMC. Its additional optimization cost and French continuation sensitivity distinguish it from the derivative-only correction (Section~\ref{sec:vbc}).

The correction's limits are equally informative. Underdispersed Canadian location marginals and selected stations' changing location spread with log scale remain beyond a translated Gaussian's reach. A separate French station-maximum diagnostic retains distributional error (Appendix~\ref{app:station-maxima}): its largest station mean is almost exactly recovered, while its joint maximization contribution remains underestimated. France also has the smallest modal support margin (Table~\ref{tab:fit-checks}) and the only observed support violations among the Gaussian reconstructions (Table~\ref{tab:draw-support}). Its residual errors are consistent with support-boundary sensitivity, without establishing a causal explanation.

The common completeness and record-length requirements improve comparability of preprocessing, but do not make the observation processes equivalent. Canadian blocks maximize monthly snowfall totals, whereas the UK and French blocks maximize daily measurements. Missing days and months remain unmodeled. The Canadian GEV is fitted to positive annual maxima without modeling snow-free years, so Canadian return levels describe years with some snowfall. The fixed common-shape, conditionally independent, time-invariant model also omits temporal trends and residual event dependence. Practitioners also use areal design quantities and regional event probabilities, which require a model for dependence between simultaneous extremes \citep{davison2012}. Our conditionally independent marginal model does not estimate that residual dependence; the regional functionals summarize parameter fields rather than areal design events. In particular, agreement among computational methods does not assess stationarity of French temperature extremes or the missingness assumptions.

This comparison does not establish discretization convergence or propagate uncertainty in the data-derived fixed offsets. Five-chain diagnostics and chain-aware MCSEs support the observed comparisons, but shared local initialization and warmup adaptation cannot exclude bias common to all chains.

For applied work, these results suggest a division of labor. When the reported quantities are regional averages, contrasts, or exceedance areas, L-MQ-TMC provides close posterior distributions at negligible extra cost, although its accuracy should be checked against MCMC on a representative fit. When results depend on weakly informed stations, joint maxima, or tails near the support boundary, full posterior sampling is affordable enough to be the default for these applications.

\FloatBarrier
\section*{Code and data availability}
Code, configuration files, and instructions for reproducing all experiments are publicly available at \url{https://github.com/infamoussoap/SpatialGEVMCMC}. The observations are available from the MIDAS Open, M\'et\'eo-France, and Environment and Climate Change Canada archives cited in Section~\ref{sec:data}.

\FloatBarrier
\bibliographystyle{plainnat}
\bibliography{references}
\FloatBarrier
\appendix
\section{Spatial variation in the fitted GEV parameters}
\label{app:parameter-maps}

\begin{figure}[htbp]
\centering
\includegraphics[width=\linewidth]{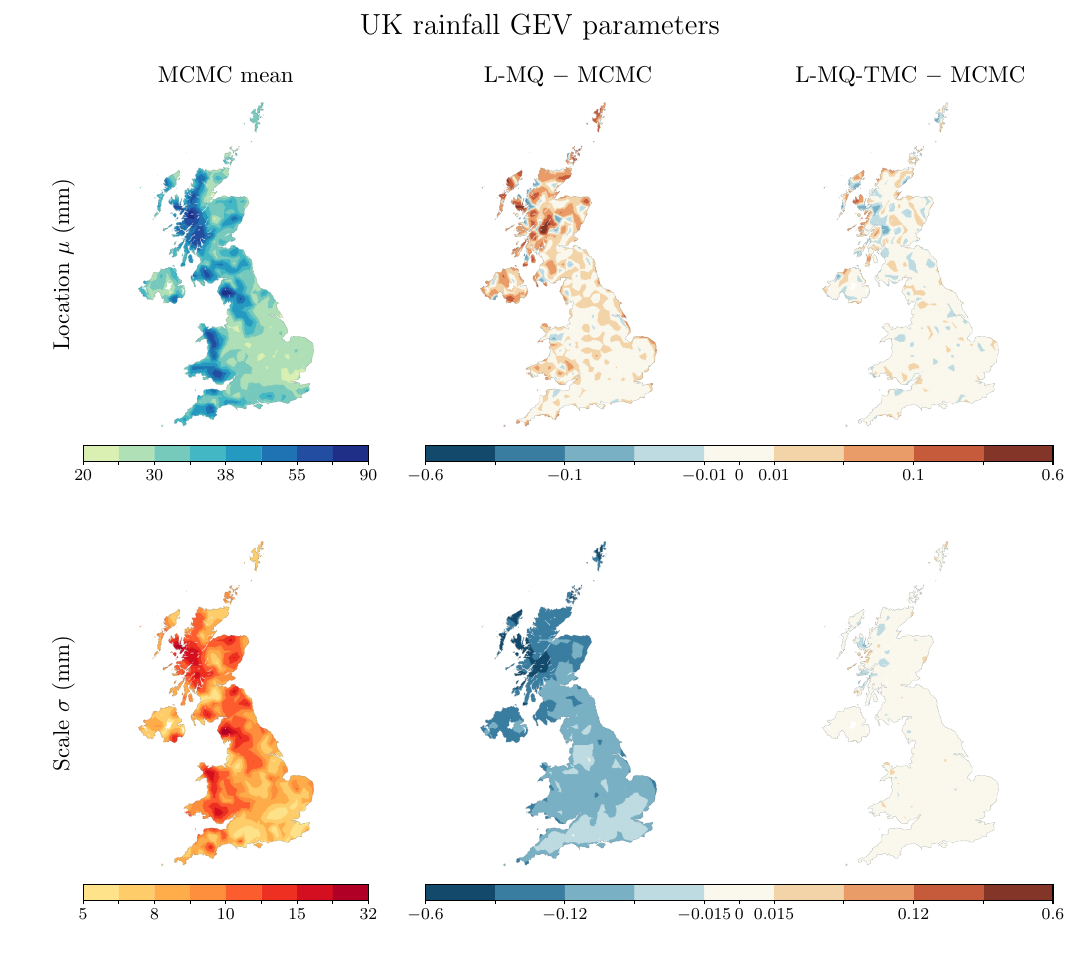}
\caption{GEV parameters for UK annual maxima of daily rainfall. Location and scale are in millimeters. The reference means use all 25,000 retained MCMC draws. The center and right columns show L-MQ minus MCMC and L-MQ-TMC minus MCMC, respectively. The MCMC mean of the common shape $\xi$ is $0.07098$, and the difference from MCMC is $-0.000058$ for both approximations.}
\label{fig:parameter-means-uk}
\end{figure}

\begin{figure}[htbp]
\centering
\includegraphics[width=\linewidth]{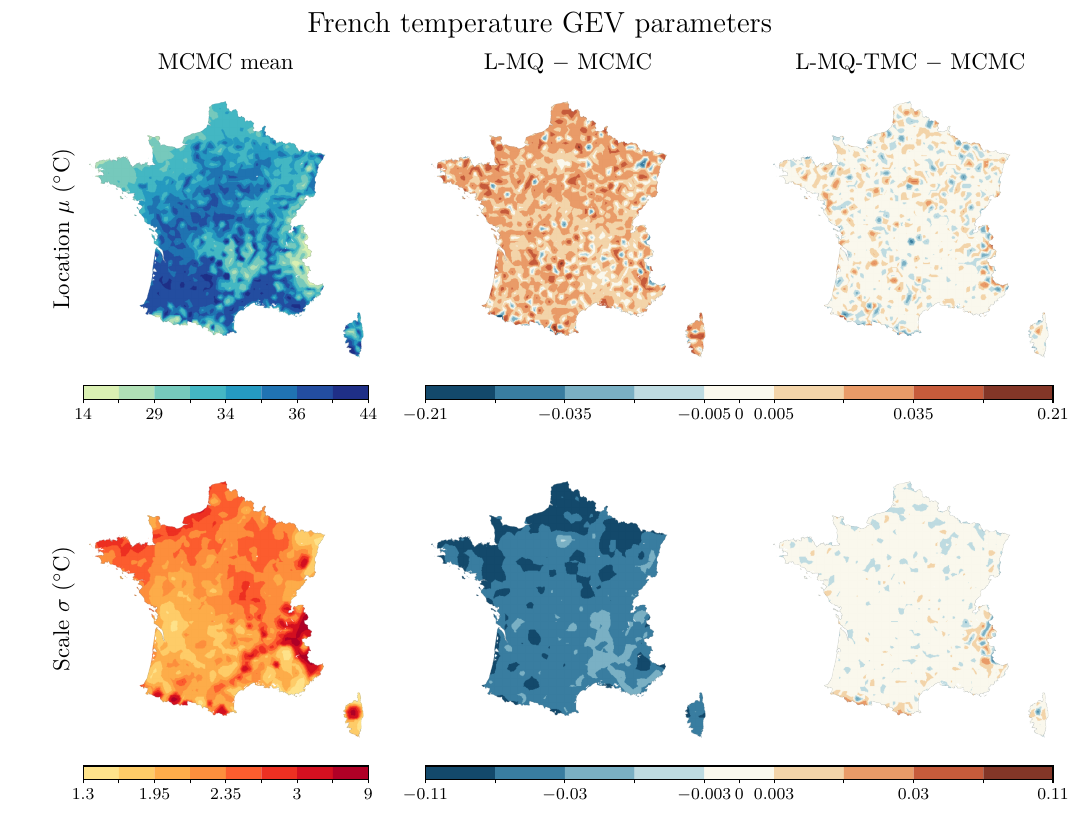}
\caption{GEV parameters for annual maxima of daily maximum temperature in metropolitan France, including Corsica. Location and scale are in degrees Celsius. The reference means use all 25,000 retained MCMC draws. The center and right columns show the signed differences from MCMC. The MCMC mean of the common shape $\xi$ is $-0.20870$, and the difference from MCMC is $-0.000237$ for both approximations.}
\label{fig:parameter-means-france}
\end{figure}

\begin{figure}[htbp]
\centering
\includegraphics[width=\linewidth]{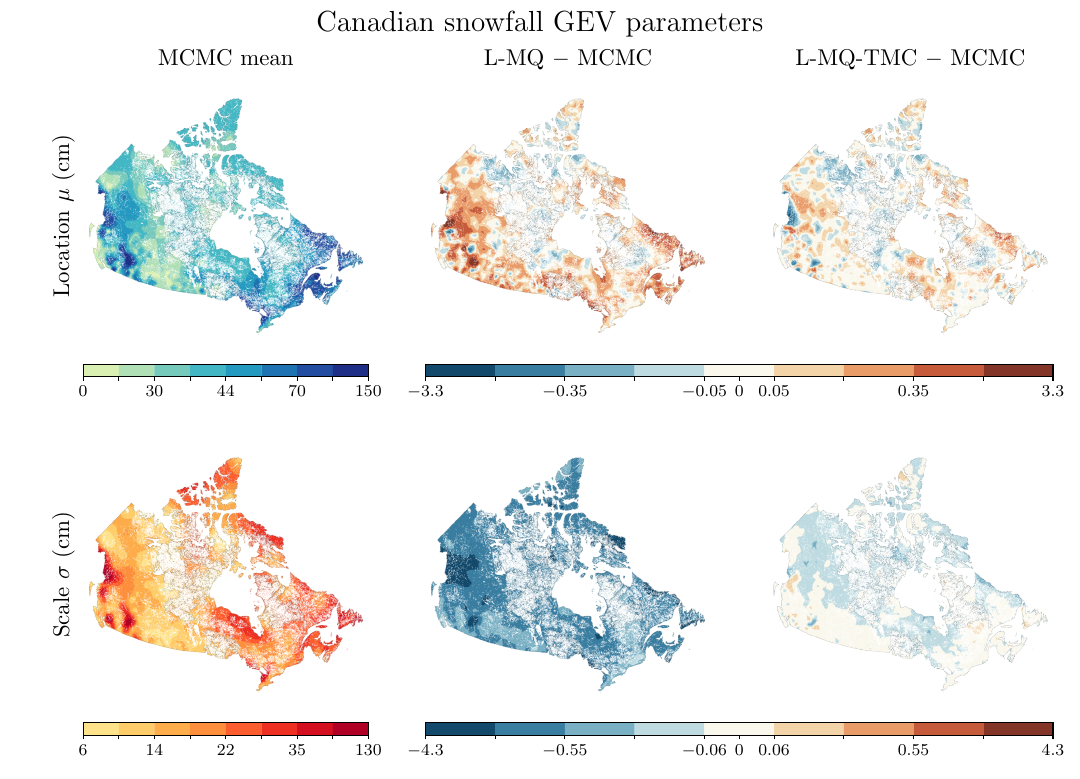}
\caption{GEV parameters for Canadian annual maxima of monthly snowfall totals. Location and scale are in centimeters. The reference means use all 65,000 retained MCMC draws. The center and right columns show the signed differences from MCMC. The MCMC mean of the common shape $\xi$ is $0.02080$, and the difference from MCMC is $-0.000373$ for both approximations.}
\label{fig:parameter-means-canada}
\end{figure}

Figures~\ref{fig:parameter-means-uk}--\ref{fig:parameter-means-canada} show the mean GEV location and scale fields for each national fit. At each mapped location $\boldsymbol s$, we average $\mu(\boldsymbol s)$ and $\sigma(\boldsymbol s)=\exp\{\eta(\boldsymbol s)\}$ over the retained MCMC draws. For scale, we interpolate log scale on the mesh and exponentiate within each draw before averaging. The MCMC scale maps describe the finite runs. At off-station locations where a component of log scale is unconstrained by the station likelihood, its exponential can have an infinite posterior mean under the fitted field-SD prior.

The two comparison columns subtract the MCMC mean from the corresponding L-MQ or L-MQ-TMC mean at each location. We evaluate the Gaussian scale means as $\exp\{m_\eta(\boldsymbol s)+v_\eta(\boldsymbol s)/2\}$, where $m_\eta$ and $v_\eta$ are the marginal mean and variance of log scale under the approximation. Blue indicates a lower mean than MCMC and orange a higher mean. The two difference maps for each parameter use the same color intervals. Each national fit has a single shape parameter $\xi$, and each caption reports its MCMC mean and the difference from MCMC for both approximations.

Across the three countries, L-MQ gives lower scale estimates than the MCMC reference throughout the displayed land grids. TMC brings the scale fields closer to the reference, although negative differences remain across parts of Canada. The two shape differences coincide in each country because TMC shifts only the location and log-scale fields and retains the L-MQ distribution of shape.

\FloatBarrier

\section{Sampler transformations and implementation details}
\label{app:sampler}
\subsection{Full modal noncentering}\label{app:transforms}
Let $\boldsymbol{M}$ be the diagonal lumped mass matrix and $\boldsymbol{G}$ the stiffness matrix for the fixed triangular mesh. Diagonalize the fixed operator once:
\begin{equation}\label{eq:app_modal_basis}
 \boldsymbol{M}^{-1/2}\boldsymbol{G}\boldsymbol{M}^{-1/2}
 =\boldsymbol{U}\diag(\lambda_j)\boldsymbol{U}^\top,
 \qquad \boldsymbol{W}=\boldsymbol{M}^{-1/2}\boldsymbol{U},\qquad
 \boldsymbol{W}^{-1}=\boldsymbol{U}^\top\boldsymbol{M}^{1/2}.
\end{equation}
Natural boundary conditions produce a constant null mode on the connected mesh. Its eigenvalue is set to its known value zero after numerical verification; the remaining modes are retained without truncation. For $\boldsymbol{h}_g=(\log\rho_g,\log\tau_g)^\top$ define
\begin{equation}\label{eq:app_noncentering}
 \omega_{gj}(\boldsymbol{h}_g)
 =\frac{\sqrt{4\pi}\,\kappa_g\tau_g}
 {\kappa_g^2+\lambda_j},\qquad
 \boldsymbol{B}_g=\boldsymbol{W}\diag(\omega_{gj}),\qquad
 \boldsymbol{v}_g=\boldsymbol{B}_g\boldsymbol{z}_g.
\end{equation}
Write $\boldsymbol{\omega}_g$ for the vector of modal SDs $\omega_{gj}$. This gives $\boldsymbol{B}_g\boldsymbol{B}_g^\top=\boldsymbol{Q}_g^{-1}$ for the mass-lumped precision in the model. Under this transformation, the Gaussian prior determinant cancels the noncentering Jacobian, yielding the field/shape target in \eqref{eq:mcmc_field_target}. The physical coefficient vector $\boldsymbol{v}_g$ is the mesh field around its fixed offset; station values still require the interpolation matrix. The common shape has a normal prior on $\zeta$ itself, so $\xi=0.5\tanh\zeta$ does not require a shape Jacobian when the density is evaluated in $\zeta$.

During the centered update, the modal physical coefficients $\boldsymbol{u}_g=\boldsymbol{W}^{-1}\boldsymbol{v}_g =\boldsymbol{\omega}_g(\boldsymbol{h}_g) \mathbin{\odot}\boldsymbol{z}_g$ remain fixed, where $\odot$ denotes elementwise multiplication. The hyperparameter conditional can be evaluated without a dense precision rebuild or observation likelihood:
\begin{equation}\label{eq:app_centered_log}
 \begin{split}
 \log\pi(\boldsymbol{h}\mid\boldsymbol{v},\zeta,\boldsymbol{y})
 &=\sum_g\left\{-\tfrac12\sum_j
 \frac{u_{gj}^2}{\omega_{gj}(\boldsymbol{h}_g)^2}
 -\sum_j\log\omega_{gj}(\boldsymbol{h}_g)
 +\log p_{h,g}(\boldsymbol{h}_g)\right\}+C,\\
 \log p_{h,g}(\boldsymbol{h}_g)
 &=-h_{g,1}+h_{g,2}-\lambda_{\rho,g}e^{-h_{g,1}}
 -\lambda_{\tau,g}e^{h_{g,2}}+C_g.
 \end{split}
\end{equation}
Here $h_{g,1}=\log\rho_g$ and $h_{g,2}=\log\tau_g$ are the scalar entries of $\boldsymbol{h}_g$. After accepting $\boldsymbol{h}'$, set $z_{gj}=u_{gj}/\omega_{gj}(\boldsymbol{h}'_g)$. This remapping leaves $\boldsymbol{v}_g$ unchanged. If the hyperparameter proposal or remapped state is nonfinite, the hyperparameter block is rejected.

\subsection{Split map and GEV support}\label{app:sampler-support}
For a fixed block reference with precision $\boldsymbol{L}\boldsymbol{L}^\top$, let $\boldsymbol{F}=\boldsymbol{L}^{-\top}$, $\boldsymbol{w}=\boldsymbol{L}^\top(\boldsymbol{q}-\boldsymbol{c})$, and $\mathcal R(\boldsymbol{w})=-\log\pi_b (\boldsymbol{c}+\boldsymbol{F}\boldsymbol{w}) -\|\boldsymbol{w}\|^2/2$. One split step of size $\epsilon=\Theta/K$ applies
\begin{align}\label{eq:app_split}
 \boldsymbol{p}&\leftarrow\boldsymbol{p}
 -\tfrac\epsilon2\nabla\mathcal R(\boldsymbol{w}),\nonumber\\
 (\boldsymbol{w},\boldsymbol{p})
 &\leftarrow(\cos\epsilon\,\boldsymbol{w}
 +\sin\epsilon\,\boldsymbol{p},
 -\sin\epsilon\,\boldsymbol{w}
 +\cos\epsilon\,\boldsymbol{p}),\\
 \boldsymbol{p}&\leftarrow\boldsymbol{p}
 -\tfrac\epsilon2\nabla\mathcal R(\boldsymbol{w}).\nonumber
\end{align}
Both expressions in the rotation use the pre-rotation pair. The force is equivalently $-\nabla\mathcal R(\boldsymbol{w}) =\boldsymbol{F}^\top\nabla\log\pi_b (\boldsymbol{c}+\boldsymbol{F}\boldsymbol{w})+\boldsymbol{w}$. For $K$ steps, merged adjacent kicks require $K+1$ force evaluations. The affine transform has constant Jacobian during a block. Symmetric position-dependent kicks and Gaussian rotations give a volume-preserving map that is reversible under momentum reflection; with refreshed Gaussian momentum, the supported-target acceptance probability \eqref{eq:mcmc_accept} corrects its integration error \citep{beskos2011,shahbaba2014,ottobre2016}.

The position-dependent force continuation preserves the kick/rotation reversibility; it does not change the likelihood used for inference.

\subsection{Reference fitting and adaptation}
References are fitted by L-BFGS-B with trust-region refinement; no curvature flooring is needed. Each refresh refits the field/shape reference at the pooled median hyperparameters, keeping the centered reference fixed; refits never alter live chain states.

Each reference window pools its latter half across chains, including repetitions after rejection. Shared block angles adapt by dual averaging toward acceptance 0.8, then freeze before the final settling period (Table~\ref{tab:sampler_tuning}).

\begin{table}[htbp]
\centering\small
\caption{Discarded adaptation schedule and frozen block angles, shared across chains and retained continuations.}
\label{tab:sampler_tuning}\label{tab:frozen-angles}
\begin{tabular}{lrrr}
\toprule
 & UK & France & Canada\\
\midrule
Reference updates & 1,000; 2,000 & 1,000; 2,000 &
 \shortstack{1,000; 2,000;\\3,000; 4,000}\\
Angle freeze iteration & 2,500 & 2,500 & 4,500\\
Split steps per block & 6 & 6 & 8\\
Frozen field/shape angle & $90^\circ$ & $90^\circ$ & $90^\circ$\\
Frozen hyperparameter angle & $90^\circ$ & $90^\circ$ & $74.955^\circ$\\
\bottomrule
\end{tabular}
\end{table}

Chains start from small perturbations (scale 0.05) of the reference center.

\subsection{Canadian retained-run checks}
\label{app:canada-continuations}
Table~\ref{tab:canada-segments} reports the Canadian diagnostic screens as the retained length increases.

\begin{table}[htbp]
\centering\small
\caption{Canada cumulative retained-run checks. Failed counts are diagnostic rows, including duplicated log and natural scales, rather than distinct model parameters. Warmup occurs only before the initial segment.}
\label{tab:canada-segments}
\begin{tabular}{rrrrrl}
\toprule
Draws/chain & $\max\rh$ & Min bulk & Min tail & Failed rows & Screen\\
\midrule
10,000 & 1.0132 & 454 & 626 & 2 & Fail\\
11,000 & 1.0135 & 509 & 719 & 2 & Fail\\
12,000 & 1.0118 & 554 & 763 & 2 & Fail\\
13,000 & 1.0082 & 583 & 791 & 0 & Pass\\
\bottomrule
\end{tabular}
\end{table}

\FloatBarrier
\subsection{Comparison with Stan NUTS}\label{app:nuts}
The Edinburgh domain uses a triangle-centroid cutoff of 111.916\,km.

\paragraph{Target and preconditioning.}
At a joint mode of the noncentered posterior, let $\boldsymbol H_{\mathrm{MAP}}$ be the Hessian of the full negative log posterior, distinct from the conditional field Hessian in \eqref{eq:laplace}. The initial dense inverse metric is
\begin{equation}\label{eq:nuts-initial-metric}
 \boldsymbol C_{\mathrm{MAP}}=\boldsymbol H_{\mathrm{MAP}}^{-1}.
\end{equation}
It includes cross-derivatives among fields, shape, and hyperparameters. The maximum absolute mode score is $2.98\times10^{-5}$, and no curvature flooring is needed. All chains start at this mode; metric and step-size adaptation remain enabled during warmup. Independent JAX evaluations agree with Stan across 36 probes, including unsupported states, with log-density error below $2\times10^{-9}$ and scaled derivative error below $10^{-12}$.

\paragraph{Sampling controls and diagnostics.}
Stan 2.32.2 via \texttt{rstan} 2.32.7 is used; apart from the dense MAP-Hessian initial metric, target acceptance, and warmup length, all sampler and adaptation controls are Stan defaults. Diagnostic extrema cover the 205 sampler coordinates and 838 scientific estimands: five natural-scale hyperparameters, three quantities at each station, and two station-network functionals. Divergences are concentrated in one chain (29 of 39); BFMI ranges from 0.66 to 0.71, and no transition reaches the tree-depth limit.

\paragraph{Matched target-acceptance runs.}
Table~\ref{tab:nuts-pilots} compares the matched target-acceptance runs. No pilot is extended or substituted for the primary comparison; their tuning costs are excluded from the single-fit timings in Table~\ref{tab:nuts-cost}.

\begin{table}[htbp]
\centering\small
\caption{Matched Edinburgh NUTS tuning runs for the design of Section~\ref{sec:design}; only target acceptance changes. Divergences count the 20,000 retained transitions, excluding warmup. Rank/ESS extrema use the saved sampler-coordinate inventory; passing that screen does not remove divergence warnings.}
\label{tab:nuts-pilots}
\begin{tabular}{rrrrrrl}
\toprule
Target & Div. & Rate (\%) & $\max\widehat R$ & Min bulk
& Min tail & Screen\\
\midrule
0.8 & 1,125 & 5.625 & 1.0179 & 191 & 39 & Fail\\
0.9 & 333 & 1.665 & 1.0057 & 2547 & 1463 & Pass\\
0.95 & 284 & 1.420 & 1.0014 & 2787 & 2266 & Pass\\
0.995 & 39 & 0.195 & 1.0016 & 3059 & 2897 & Pass\\
\bottomrule
\end{tabular}
\end{table}

\paragraph{Posterior summaries and Monte Carlo uncertainty.}
Station and network quantities contribute means, SDs, and equal-tailed 95\% interval endpoints; the five hyperparameters contribute medians and interval endpoints, giving 3,347 comparisons across 838 estimands. For a summary $a$, the reported discrepancy is
\begin{equation}\label{eq:nuts-standardized-difference}
 z_a^{\mathrm{MC}}
 =\frac{\widehat a_{\mathrm{LR}}-\widehat a_{\mathrm{NUTS}}}
 {\sqrt{\widehat{\mathrm{MCSE}}_{\mathrm{LR},a}^{\,2}
       +\widehat{\mathrm{MCSE}}_{\mathrm{NUTS},a}^{\,2}}}.
\end{equation}
Mean MCSEs use mean-specific ESS. SD MCSEs use the autocorrelation of squared deviations and the delta method; endpoint MCSEs use quantile-specific ESS and the order-statistic method. The pointwise $|z_a^{\mathrm{MC}}|\leq1.96$ screen is supplemented by Holm adjustment of two-sided normal-approximation $p$-values across this comparison family. This adjustment accommodates dependence between stations but retains the approximation and MCSE assumptions.

\begin{figure}[htbp]
\centering
\includegraphics[width=\linewidth]
{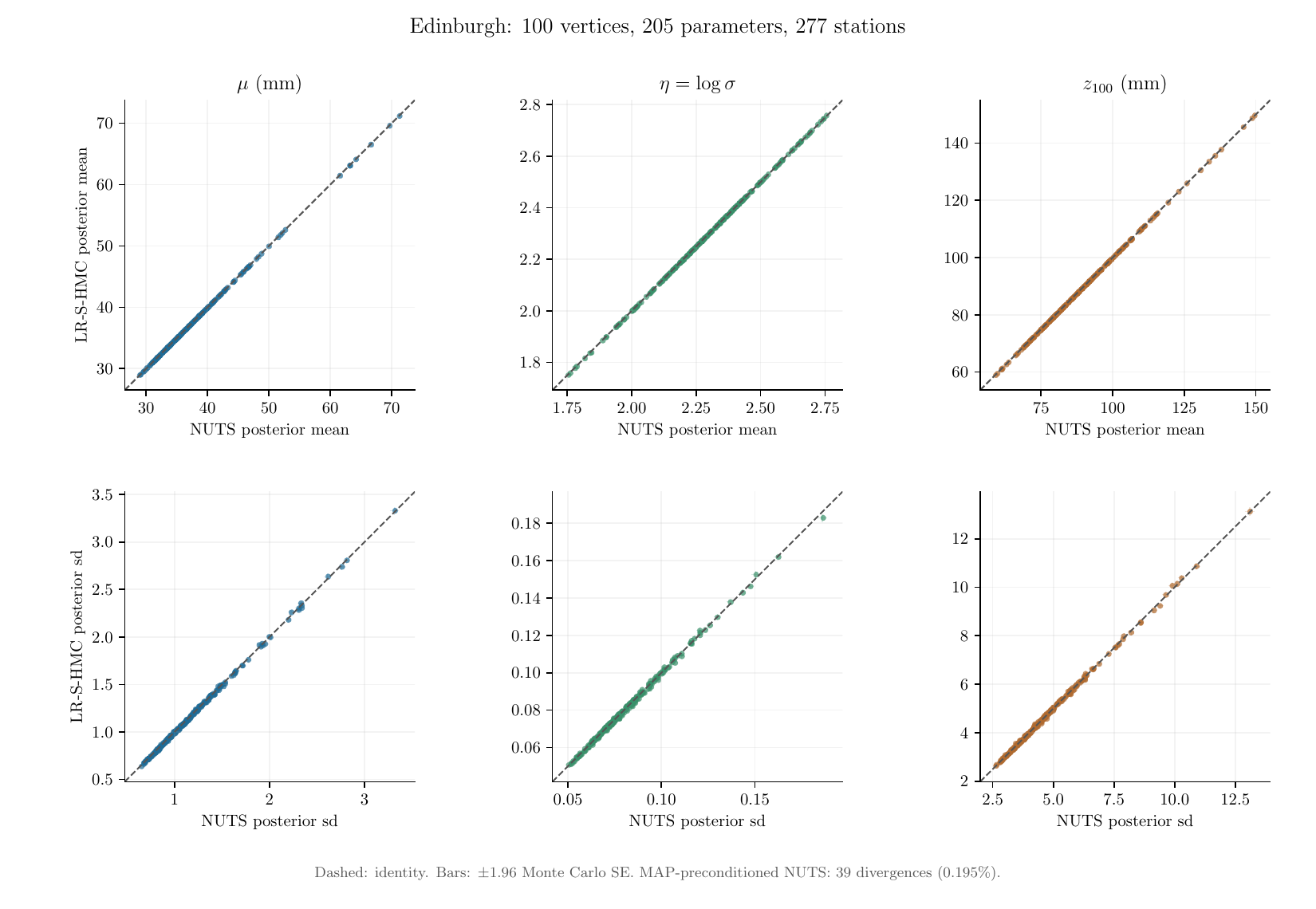}
\caption{LR-S-HMC and NUTS posterior means and SDs for station location, log scale, and 100-year return levels on the local UK domain. Each point is a station; dashed lines mark equality. Bars are $\pm1.96$ MCSE for each plotted summary, rather than posterior credible intervals. Equal axis limits aid comparison.}
\label{fig:nuts-posterior-agreement}
\end{figure}

Table~\ref{tab:nuts-hyperparameters} reports the hyperparameter summaries, and Figure~\ref{fig:nuts-posterior-agreement} compares station means and SDs.

\begin{table}[htbp]
\centering\small
\setlength{\tabcolsep}{4pt}
\caption{ Natural-scale hyperparameter posterior medians and quantiles from the four-chain comparison. Ranges are in km, $\tau_\mu$ is in mm, and $\tau_\eta$ and $\xi$ are dimensionless. The interval endpoints are posterior percentiles. For each method pair, the final column gives the maximum absolute standardized difference across the median and two endpoints from Eq.~\eqref{eq:nuts-standardized-difference}, using the saved primary MCSE estimates. }
\label{tab:nuts-hyperparameters}
\begin{tabular*}{\linewidth}
{@{\extracolsep{\fill}}llrrrr@{}}
\toprule
Parameter & Method & Median & $q_{.025}$ & $q_{.975}$ & Max. $|z|$\\
\midrule
$\rho_\mu$ & LR-S-HMC & 69.96 & 43.98 & 137.30 & 0.826\\
 & NUTS & 69.68 & 43.88 & 134.19 & \\
$\tau_\mu$ & LR-S-HMC & 8.33 & 6.52 & 13.23 & 1.005\\
 & NUTS & 8.32 & 6.50 & 13.17 & \\
$\rho_\eta$ & LR-S-HMC & 53.84 & 31.33 & 110.45 & 0.901\\
 & NUTS & 54.18 & 31.07 & 110.20 & \\
$\tau_\eta$ & LR-S-HMC & 0.2799 & 0.2164 & 0.4211 & 0.924\\
 & NUTS & 0.2799 & 0.2156 & 0.4144 & \\
$\xi$ & LR-S-HMC & 0.03950 & 0.01956 & 0.06034 & 1.054\\
 & NUTS & 0.03956 & 0.01928 & 0.05996 & \\
\bottomrule
\end{tabular*}
\end{table}

Table~\ref{tab:nuts-sensitivity} checks sensitivity to alternative batch-based MCSE estimates.

\begin{table}[htbp]
\centering\small
\setlength{\tabcolsep}{4pt}
\caption{ Batch-MCSE sensitivity over 3,347 paired comparisons at each batch count: 833 station or network quantities contribute means, SDs, and interval endpoints, while five hyperparameters contribute medians and 95\% interval endpoints. Standardized differences divide LR-S-HMC minus NUTS by the root-sum-square batch MCSE. The signed maximum and its globally Holm-adjusted p-value concern the comparison attaining the largest absolute discrepancy at each batch count. Holm adjustment uses comparisons with available p-values and significance level 0.05. Absence of rejections does not establish posterior equivalence. }
\label{tab:nuts-sensitivity}
\begin{tabular*}{\linewidth}
{@{\extracolsep{\fill}}rrrrrr@{}}
\toprule
\shortstack{Batches\\per chain} & \shortstack{Overall\\max. $|z|$} & \shortstack{Hyperparameter\\max. $|z|$} & \shortstack{Signed\\maximum $z$} & \shortstack{Maximum\\Holm $p$} & \shortstack{Global Holm\\rejections}\\
\midrule
20 & 3.522 & 1.124 & +3.522 & 1.000 & 0\\
40 & 3.694 & 1.152 & +3.694 & 0.738 & 0\\
80 & 3.642 & 1.149 & -3.642 & 0.905 & 0\\
\bottomrule
\end{tabular*}
\end{table}

\paragraph{Observed efficiency.}
Table~\ref{tab:nuts-efficiency} gives group median ESS per whole-job wall-clock second.

\begin{table}[htbp]
\centering\small
\setlength{\tabcolsep}{4pt}
\caption{ Median bulk and tail ESS per second within each reported natural-scale or derived-quantity group; $n$ is the number of quantities. Ratios are LR-S-HMC divided by NUTS, using whole-job elapsed time. Ratios using retained-phase sampling time differ by at most 0.1 at the reported precision. These are observed efficiency summaries, not a benchmark at established matched posterior accuracy. }
\label{tab:nuts-efficiency}
\begin{tabular*}{\linewidth}
{@{\extracolsep{\fill}}lrrrrrrr@{}}
\toprule
Group & $n$ & \shortstack{LR-S-HMC\\bulk} & \shortstack{LR-S-HMC\\tail} & \shortstack{NUTS\\bulk} & \shortstack{NUTS\\tail} & \shortstack{Bulk\\ratio} & \shortstack{Tail\\ratio}\\
\midrule
Hyperparameters & 5 & 83.37 & 79.61 & 2.21 & 2.76 & 37.8 & 28.9\\
Station $\mu$ & 277 & 174.90 & 180.29 & 15.85 & 8.91 & 11.0 & 20.2\\
Station $\eta$ & 277 & 175.84 & 182.41 & 15.69 & 9.00 & 11.2 & 20.3\\
Station $z_{100}$ & 277 & 180.41 & 182.43 & 15.77 & 8.88 & 11.4 & 20.5\\
Network functionals & 2 & 180.57 & 169.25 & 14.63 & 8.85 & 12.3 & 19.1\\
\bottomrule
\end{tabular*}
\end{table}
\section{Sparse derivation and numerical validation}\label{app:numerics}
\subsection{Local moment expansion}\label{app:t3-derivation}
Let $\boldsymbol{e}=\boldsymbol{v}-\widehat{\boldsymbol{v}}$ and expand the conditional negative log posterior at an interior mode:
\begin{equation}\label{eq:app_t3_expansion}
 \Phi(\widehat{\boldsymbol{v}}+\boldsymbol{e},\boldsymbol{\theta}^\star)
 =\Phi(\widehat{\boldsymbol{v}},\boldsymbol{\theta}^\star)
 +\tfrac12\boldsymbol{e}^\top\boldsymbol{H}\boldsymbol{e}
 +\tfrac16\sum_{a,b,c}T_{abc}e_ae_be_c+\cdots.
\end{equation}
Write $\E_0$ for expectation under $\N(\boldsymbol{0}_{2m},\boldsymbol{\Sigma})$, $\boldsymbol{\Sigma}=\boldsymbol{H}^{-1}$. Expanding the numerator and denominator of the posterior mean integral retains the leading term
\begin{equation}\label{eq:app_t3_fourth_moments}
 \E[e_j\mid\boldsymbol{y},\boldsymbol{\theta}^\star]
 \simeq-\tfrac16\sum_{a,b,c}T_{abc}\E_0[e_je_ae_be_c].
\end{equation}
The Gaussian numerator has zero mean, and the first denominator perturbation is an odd Gaussian moment. The fourth-moment identity is
\begin{equation}\label{eq:app_t3_wick}
 \E_0[e_je_ae_be_c]
 =\Sigma_{ja}\Sigma_{bc}+\Sigma_{jb}\Sigma_{ac}
 +\Sigma_{jc}\Sigma_{ab}.
\end{equation}
Symmetry of the tensor with entries $T_{abc}$ makes the three contractions equal, yielding Equation~\eqref{eq:t3_shift}. This is the classical leading standard Laplace moment expansion; related fully exponential moment methods are developed by \citet{tierney1986,tierney1989}. We apply this local term jointly to the two spatial fields, retaining the L-MQ reconstruction for joint posterior targets.

This derivation is a local expansion of integrals. It does not define a global density by exponentiating a cubic polynomial, which need not be integrable. Its usual asymptotic interpretation requires posterior concentration around a smooth interior mode with sufficiently controlled remainders. Increasing mesh dimension and heterogeneous short records do not automatically satisfy such conditions. High-dimensional Laplace accuracy depends on how information grows with the integrated dimension, rather than on total sample size alone \citep{shun1995,ogden2021}. These results do not provide an error bound for the present spatial GEV expansion. We consequently report empirical reference accuracy without claiming a uniform asymptotic error order for these fitted spatial models.

\subsection{Station contractions and sparse solves}
The field prior is Gaussian with precision $\boldsymbol{Q}=\diag(\boldsymbol{Q}_\mu,\boldsymbol{Q}_\eta)$, so its third derivative is zero. Let $\boldsymbol{\psi}_i=(\mu_i,\eta_i)^\top$ and $\boldsymbol{P}_i=\diag(\boldsymbol{a}_i^\top, \boldsymbol{a}_i^\top)$. Station predictors are affine in the fields; the fixed offsets vanish on differentiation. With station indices $1,2$ corresponding to $\mu,\eta$, the chain rule gives
\begin{equation}\label{eq:app_t3_chain_rule}
 T_{abc}=\sum_i\sum_{r,s,t=1}^2
 P_{i,ra}P_{i,sb}P_{i,tc}T_{i,rst},\qquad
 \boldsymbol{C}_i=\boldsymbol{P}_i\boldsymbol{\Sigma}\boldsymbol{P}_i^\top.
\end{equation}
Contracting over the last two field indices therefore requires only $\boldsymbol{C}_i$. With symmetric mixed derivatives, the two components of $\boldsymbol{t}_i$ are explicitly
\begin{align}\label{eq:app_t3_components}
 t_{i,\mu}
 &=T_{i,\mu\mu\mu}C_{i,\mu\mu}
 +2T_{i,\mu\mu\eta}C_{i,\mu\eta}
 +T_{i,\mu\eta\eta}C_{i,\eta\eta},\\
 t_{i,\eta}
 &=T_{i,\mu\mu\eta}C_{i,\mu\mu}
 +2T_{i,\mu\eta\eta}C_{i,\mu\eta}
 +T_{i,\eta\eta\eta}C_{i,\eta\eta}.
\end{align}
The factor two accounts for the two mixed covariance terms. Sparse interpolation assembles $\boldsymbol{b}=\sum_i\boldsymbol{P}_i^\top \boldsymbol{t}_i$, followed by the single solve in Equation~\eqref{eq:t3_sparse_solve}. Selected inverse entries come from Takahashi recursion on the existing factor, so no field-sized inverse, third-order tensor, or new factorization is formed.

\subsection{L-MQ sensitivity and analytic target means}
For the original conditional mode, differentiating its stationarity condition gives
\begin{equation}\label{eq:app_mode_sensitivity}
 \boldsymbol{J}=-\boldsymbol{H}^{-1}\boldsymbol{H}_{v\theta},\qquad
 \boldsymbol{H}_{v\theta}
 =\nabla_{\boldsymbol{v}}\nabla_{\boldsymbol{\theta}}^\top\Phi
 (\widehat{\boldsymbol{v}},\boldsymbol{\theta}^\star).
\end{equation}
Sparse solves evaluate this sensitivity. The Laplace hyperparameter gradient includes total derivatives of the field log determinant, $\mathrm d\log|\boldsymbol{H}|=\operatorname{tr} (\boldsymbol{H}^{-1}\mathrm d\boldsymbol{H})$, including mode dependence. Selected inverse entries provide these traces without stochastic estimation. Two central-difference calculations check the five-dimensional marginal curvature, with agreement reported in Table~\ref{tab:fit-checks}.

The conditional covariance used in the third-order contraction differs from the final marginal station covariance:
\begin{equation}\label{eq:app_marginal_station_covariance}
 \boldsymbol{C}_i^{\text{L-MQ}}=\boldsymbol{C}_i
 +\boldsymbol{P}_i\boldsymbol{J}\boldsymbol{V}\boldsymbol{J}^\top
 \boldsymbol{P}_i^\top.
\end{equation}
The correction uses $\boldsymbol{C}_i$ because it is derived at fixed $\boldsymbol{\theta}^\star$. 

For analytic return-level means, let $m_{\eta,i}$ and $s_{\eta,i}^2$ be the L-MQ marginal log-scale mean and variance. Let $\zeta\sim\N(m_\zeta,s_\zeta^2)$ denote the shape coordinate, and $c_{\eta\zeta,i}=\Cov_{\text{L-MQ}}(\eta_i,\zeta)$. Gaussian exponential tilting gives
\begin{equation}\label{eq:app_t3_tilting}
 \gamma_i=e^{m_{\eta,i}+s_{\eta,i}^2/2}
 \E_{\widetilde{\zeta}\sim\N(m_\zeta+c_{\eta\zeta,i},s_\zeta^2)}
 [c_{100}\{0.5\tanh\widetilde{\zeta}\}].
\end{equation}
This is a one-dimensional Gaussian integral, evaluated with the saved L-MQ moment calculation. Its use for posterior summaries is separate from the derivative-only mean correction. Substitution into Equation~\eqref{eq:analytic_mean} gives corrected station means.

\subsection{Derivative and factorization checks}\label{app:t3-validation}
Quadratic and coupled-cubic tests verify zero correction and the negative one-half mixed-derivative contraction. All marginal optimizers converge, with numerical checks in Table~\ref{tab:fit-checks}.

\begin{table}[htbp]
\centering\small
\caption{Numerical fit checks. Gradient is the marginal objective gradient norm; curvature compares two finite-difference step sizes. Factor error is the conditional precision factorization residual. Support is the minimum $1+\xi(y-\mu)e^{-\eta}$ at the conditional mode. Final regularization is zero for all fits.}
\label{tab:fit-checks}
\begin{tabular}{lrrrr}
\toprule
Dataset & Gradient & Curvature & Factor error & Support\\
\midrule
UK & 1.49e-04 & 2.18e-06 & 4.11e-16 & 0.846\\
France & 1.05e-04 & 2.23e-06 & 3.37e-16 & 0.138\\
Canada & 6.35e-05 & 1.00e-05 & 3.18e-16 & 0.948\\
\bottomrule
\end{tabular}
\end{table}

\subsection{Scope of matched implementation validation}\label{app:validation}
Synthetic checks compare the JAX Gaussian construction with an aligned TMB implementation using SpatialGEV likelihood and prior utilities. Alignment requires identical observations, finite-element matrices, fixed offsets, field scaling, PC priors, and the shape link $\xi=0.5\tanh\zeta$, $\zeta\sim\N(0,0.65^2)$. Without this alignment the upstream template can specify a different posterior. Agreement between the aligned implementations checks the Gaussian construction, rather than its posterior accuracy.

\subsection{VBC implementation}\label{app:vbc-implementation}
The mean objective uses a finite $C^2$ continuation of the log-support term below $b=0.02$ with a quartic penalty, and quadratic continuation of exponentials above 20. Tensor Gauss--Hermite quadrature uses 17, 25, then 33 nodes per predictor coordinate. Sparse Newton optimization starts at zero and at the TMC shift; the lower converged objective initializes the quadrature refinements.

\section{Additional posterior comparisons}\label{app:posteriors}
\subsection{Full reporting partitions}\label{app:full-partitions}
Table~\ref{tab:regions} combines scalar summaries and finite-pool benchmarks for the full reporting partitions and detailed contrasts.

\begingroup
\footnotesize
\setlength{\tabcolsep}{3pt}
\begin{longtable}{@{}>{\raggedright\arraybackslash}p{.29\linewidth}rrrrr@{}}
\caption{Area-averaged local 100-year return levels for the full partitions and the eight detailed paired contrasts. MCMC entries are medians and central 95\% credible intervals on the mapped-land domains (Appendix~\ref{app:area-numerics}). The final column gives the central 95\% finite-pool noise range for $K$, using the benchmark of Section~\ref{sec:accuracy}.}
\label{tab:regions}\label{tab:areal-noise}\\
\toprule
Region or contrast & MCMC median [95\% CrI] & $K_{\mathrm{MQ}}$ & $K_{\mathrm{TMC}}$ & $K_{\mathrm{VBC}}$ & Benchmark \\
\midrule
\endfirsthead
\multicolumn{6}{l}{\tablename\ \thetable\ (continued)}\\
\toprule
Region or contrast & MCMC median [95\% CrI] & $K_{\mathrm{MQ}}$ & $K_{\mathrm{TMC}}$ & $K_{\mathrm{VBC}}$ & Benchmark \\
\midrule
\endhead
\bottomrule
\endfoot
\multicolumn{6}{l}{\textbf{UK: area averages (mm)}} \\
Scotland & 95.93 [94.47, 97.49] & 0.473 & 0.012 & 0.016 & $[0.005,0.015]$ \\
England & 79.78 [79.06, 80.52] & 0.335 & 0.013 & 0.013 & $[0.005,0.016]$ \\
Wales & 102.73 [101.14, 104.37] & 0.220 & 0.009 & 0.010 & $[0.005,0.015]$ \\
Northern Ireland & 76.85 [74.67, 79.18] & 0.277 & 0.013 & 0.015 & $[0.005,0.016]$ \\
\multicolumn{6}{l}{\textbf{UK: paired contrasts (mm)}} \\
Scotland - England & 16.15 [14.77, 17.58] & 0.346 & 0.011 & 0.015 & $[0.005,0.014]$ \\
Scotland - Wales & -6.80 [-8.72, -4.85] & 0.202 & 0.011 & 0.014 & $[0.005,0.014]$ \\
Scotland - Northern Ireland & 19.09 [16.49, 21.62] & 0.050 & 0.008 & 0.008 & $[0.005,0.014]$ \\
England - Wales & -22.95 [-24.45, -21.48] & 0.071 & 0.008 & 0.008 & $[0.004,0.014]$ \\
England - Northern Ireland & 2.93 [0.64, 5.12] & 0.176 & 0.010 & 0.012 & $[0.005,0.016]$ \\
Wales - Northern Ireland & 25.88 [23.20, 28.48] & 0.108 & 0.010 & 0.011 & $[0.005,0.015]$ \\
\midrule
\multicolumn{6}{l}{\textbf{France: area averages ($^{\circ}$C)}} \\
\^Ile-de-France & 41.61 [41.35, 41.88] & 0.313 & 0.007 & 0.009 & $[0.005,0.014]$ \\
Centre-Val de Loire & 41.93 [41.75, 42.12] & 0.513 & 0.015 & 0.014 & $[0.005,0.015]$ \\
Bourgogne-Franche-Comt\'e & 41.05 [40.87, 41.23] & 0.536 & 0.013 & 0.011 & $[0.005,0.013]$ \\
Normandie & 39.96 [39.70, 40.24] & 0.450 & 0.020 & 0.018 & $[0.005,0.014]$ \\
Hauts-de-France & 40.32 [40.01, 40.62] & 0.466 & 0.019 & 0.023 & $[0.005,0.016]$ \\
Grand Est & 40.71 [40.50, 40.93] & 0.526 & 0.021 & 0.015 & $[0.005,0.015]$ \\
Pays de la Loire & 41.03 [40.78, 41.29] & 0.440 & 0.012 & 0.005 & $[0.005,0.016]$ \\
Bretagne & 38.99 [38.71, 39.28] & 0.442 & 0.012 & 0.027 & $[0.005,0.015]$ \\
Nouvelle-Aquitaine & 41.64 [41.50, 41.78] & 0.633 & 0.020 & 0.014 & $[0.005,0.014]$ \\
Occitanie & 41.71 [41.58, 41.85] & 0.584 & 0.014 & 0.009 & $[0.005,0.015]$ \\
Auvergne-Rh\^one-Alpes & 40.74 [40.60, 40.87] & 0.499 & 0.018 & 0.018 & $[0.005,0.014]$ \\
Provence-Alpes-C\^ote d'Azur & 40.39 [40.20, 40.59] & 0.422 & 0.008 & 0.036 & $[0.005,0.015]$ \\
Corse & 41.44 [41.05, 41.84] & 0.237 & 0.006 & 0.006 & $[0.005,0.015]$ \\
\multicolumn{6}{l}{\textbf{France: paired contrasts ($^{\circ}$C)}} \\
Hauts-de-France - Provence-Alpes-C\^ote d'Azur & -0.08 [-0.43, 0.28] & 0.176 & 0.015 & 0.007 & $[0.004,0.015]$ \\
\midrule
\multicolumn{6}{l}{\textbf{Canada: area averages (cm)}} \\
Canada's North & 116.01 [101.25, 134.46] & 0.227 & 0.031 & 0.035 & $[0.004,0.016]$ \\
British Columbia & 174.78 [161.24, 193.00] & 0.278 & 0.017 & 0.021 & $[0.004,0.013]$ \\
Prairies & 101.27 [91.58, 115.08] & 0.242 & 0.028 & 0.032 & $[0.004,0.014]$ \\
Ontario & 152.28 [126.16, 199.08] & 0.100 & 0.011 & 0.012 & $[0.004,0.012]$ \\
Quebec & 155.67 [129.06, 194.86] & 0.079 & 0.010 & 0.010 & $[0.004,0.014]$ \\
Atlantic & 197.68 [176.07, 230.51] & 0.151 & 0.015 & 0.016 & $[0.004,0.014]$ \\
\multicolumn{6}{l}{\textbf{Canada: paired contrasts (cm)}} \\
Atlantic - Prairies & 96.18 [70.89, 130.15] & 0.044 & 0.006 & 0.006 & $[0.004,0.012]$ \\
\end{longtable}
\endgroup

\subsection{Areal integration and functional diagnostics}
\label{app:area-numerics}
Mapped-land domains use ONS UK-country, French administrative-region, and Statistics Canada province/territory boundaries, with GSHHG coastal and inland-water masks and additional Canadian CanVec water polygons. Canadian mapped-land areas are within 1.1\% of census land areas; source versions are recorded in the repository.

Physical-area weights correct native projected area by the inverse projection areal scale. The fields remain affine in native mesh coordinates. On clipped subtriangles we integrate the location term and the exponential log-scale term, with an affine positive area Jacobian. Convexity brackets exceedance area between affine tangent and chord level-set integrals. Adaptive subdivision gives upper--lower width at most 0.001 for every retained draw, detailed region, and evaluated threshold. We save the bracket midpoint, whose error is at most 0.0005 conditional on this area measure. 

The fifteen additional region averages use the same P1 integrals, and the Canadian quadrature-sensitivity check leaves all reported table entries and CDF discrepancies unchanged.

The 201 equally spaced threshold values span 0--400\,mm for UK daily rainfall, 25--55\,\textdegree C for French temperature, and 0--1000\,cm for Canadian monthly snowfall. We retain every grid coordinate in the saved results, with the central-$p$ window used only for display.

All 23 regional averages and eight detailed paired contrasts pass the prescribed rank-$\widehat R$ and bulk/tail ESS screens. On the complete 201-threshold grids, 439/804 UK exceedance coordinates have estimable diagnostics: 434 pass and five rare-tail England coordinates at 320--328\,mm fail. Another 365 are numerically constant in the retained pool and have undefined diagnostics. France has 375/402 estimable coordinates, all passing, and 27 numerically constant coordinates; all 402 Canadian coordinates pass. Numerically constant means retained range at most $10^{-12}$, not posterior certainty. These flags are retained in the results rather than hidden by the central-$p$ display restriction.

\subsection{Reference precision}\label{app:reference-precision}
\begin{table}[htbp]
\centering\small
\caption{Reference field-mean MCSEs in matching station posterior SD units, using the larger of within-chain batch and between-chain mean standard errors.}
\label{tab:precision}
\begin{tabular}{llrrr}
\toprule
Dataset & Field & Median & 95th percentile & Maximum\\
\midrule
UK & $\mu$ & 0.00746 & 0.01143 & 0.01752\\
UK & $\eta$ & 0.00738 & 0.01091 & 0.01658\\
France & $\mu$ & 0.00804 & 0.01170 & 0.01566\\
France & $\eta$ & 0.00801 & 0.01205 & 0.01830\\
Canada & $\mu$ & 0.00505 & 0.00830 & 0.01664\\
Canada & $\eta$ & 0.00519 & 0.00924 & 0.01895\\
\bottomrule
\end{tabular}
\end{table}

\subsection{Monte Carlo benchmarks and interval contents for areal targets}
\label{app:areal-precision}
Table~\ref{tab:regions} gives each target's noise range using the benchmark of Section~\ref{sec:accuracy}. Table~\ref{tab:areal-tails} gives its two strict interval tail contents; neither calculation requires finite posterior moments.

\begin{table}[p]
\centering\small
\caption{Reference tail contents outside the empirical central 95\% intervals of the areal approximations. $L=\Pr_{\mathrm{MCMC}}(Y<q_{.025})$ and $U=\Pr_{\mathrm{MCMC}}(Y>q_{.975})$ use strict inequalities and inverse-empirical-CDF quantiles; unlike $D$ and $S$, these quantities require no finite posterior moments.}
\label{tab:areal-tails}
\begin{tabular}{lrrrrrr}
\toprule
& \multicolumn{2}{c}{L-MQ} & \multicolumn{2}{c}{L-MQ-TMC} & \multicolumn{2}{c}{L-MQ-VBC} \\
Region or contrast & $L$ & $U$ & $L$ & $U$ & $L$ & $U$ \\
\midrule
\multicolumn{7}{l}{\textbf{UK: area averages}} \\
Scotland & 0.000 & 0.243 & 0.024 & 0.029 & 0.023 & 0.029 \\
England & 0.003 & 0.141 & 0.025 & 0.027 & 0.025 & 0.027 \\
Wales & 0.007 & 0.086 & 0.027 & 0.028 & 0.027 & 0.028 \\
Northern Ireland & 0.005 & 0.109 & 0.026 & 0.028 & 0.026 & 0.028 \\
\midrule
\multicolumn{7}{l}{\textbf{UK: paired contrasts}} \\
Scotland - England & 0.002 & 0.138 & 0.023 & 0.025 & 0.022 & 0.026 \\
Scotland - Wales & 0.008 & 0.082 & 0.026 & 0.027 & 0.025 & 0.028 \\
Scotland - Northern Ireland & 0.023 & 0.038 & 0.027 & 0.027 & 0.027 & 0.027 \\
England - Wales & 0.041 & 0.017 & 0.027 & 0.025 & 0.027 & 0.025 \\
England - Northern Ireland & 0.068 & 0.010 & 0.027 & 0.027 & 0.027 & 0.027 \\
Wales - Northern Ireland & 0.050 & 0.016 & 0.028 & 0.027 & 0.028 & 0.027 \\
\midrule
\multicolumn{7}{l}{\textbf{France: area averages}} \\
\^Ile-de-France & 0.002 & 0.132 & 0.026 & 0.027 & 0.026 & 0.027 \\
Centre-Val de Loire & 0.000 & 0.290 & 0.024 & 0.030 & 0.024 & 0.029 \\
Bourgogne-Franche-Comt\'e & 0.000 & 0.323 & 0.024 & 0.032 & 0.024 & 0.032 \\
Normandie & 0.000 & 0.236 & 0.021 & 0.033 & 0.021 & 0.032 \\
Hauts-de-France & 0.001 & 0.243 & 0.025 & 0.029 & 0.030 & 0.025 \\
Grand Est & 0.001 & 0.300 & 0.024 & 0.032 & 0.024 & 0.031 \\
Pays de la Loire & 0.001 & 0.227 & 0.024 & 0.028 & 0.025 & 0.027 \\
Bretagne & 0.001 & 0.224 & 0.024 & 0.032 & 0.028 & 0.027 \\
Nouvelle-Aquitaine & 0.000 & 0.446 & 0.025 & 0.028 & 0.026 & 0.027 \\
Occitanie & 0.000 & 0.370 & 0.021 & 0.028 & 0.022 & 0.027 \\
Auvergne-Rh\^one-Alpes & 0.000 & 0.265 & 0.023 & 0.029 & 0.022 & 0.029 \\
Provence-Alpes-C\^ote d'Azur & 0.001 & 0.194 & 0.023 & 0.026 & 0.027 & 0.021 \\
Corse & 0.005 & 0.088 & 0.026 & 0.024 & 0.026 & 0.024 \\
\midrule
\multicolumn{7}{l}{\textbf{France: paired contrasts}} \\
Hauts-de-France - Provence-Alpes-C\^ote d'Azur & 0.010 & 0.069 & 0.025 & 0.028 & 0.026 & 0.026 \\
\midrule
\multicolumn{7}{l}{\textbf{Canada: area averages}} \\
Canada's North & 0.006 & 0.073 & 0.024 & 0.027 & 0.023 & 0.027 \\
British Columbia & 0.003 & 0.104 & 0.024 & 0.033 & 0.024 & 0.033 \\
Prairies & 0.005 & 0.068 & 0.025 & 0.029 & 0.024 & 0.029 \\
Ontario & 0.014 & 0.040 & 0.027 & 0.028 & 0.026 & 0.028 \\
Quebec & 0.016 & 0.036 & 0.025 & 0.026 & 0.025 & 0.026 \\
Atlantic & 0.009 & 0.049 & 0.025 & 0.027 & 0.025 & 0.027 \\
\midrule
\multicolumn{7}{l}{\textbf{Canada: paired contrasts}} \\
Atlantic - Prairies & 0.023 & 0.036 & 0.026 & 0.027 & 0.027 & 0.027 \\
\bottomrule
\end{tabular}
\end{table}

\subsection{Joint station maxima}\label{app:station-maxima}
For a fixed station set $R$, let $Z_i=z_{100}(\boldsymbol s_i)$ and $Z_R^{\max}=\max_{i\in R}Z_i$. UK North contains 296 stations with projected northing at least 700\,km; French and Canadian North are their northern latitude-count thirds (850 and 273 stations), with stable station order breaking latitude ties. Co-located stations may share reconstructed parameters; exact maximum ties receive equal attainment weights without changing values or draws. Table~\ref{tab:maximum-uplift} shows that recovering the largest station mean can leave error in the maximization contribution: the French maximum remains underdispersed and fails the upper-tail screen. This contribution depends on marginal tails and spreads as well as dependence, so it does not isolate dependence error.

\begin{table}[htbp]
\centering\small
\caption{Northern station-maximum diagnostics. $K$ is the maximum CDF discrepancy, $S$ the empirical SD ratio, and $U$ the reference mass above the approximate 95\% interval's upper endpoint. The final two columns give $\max_i\widehat\E[Z_i]$ and the maximization contribution $\widehat\E[\max_iZ_i]-\max_i\widehat\E[Z_i]$ (mm, degrees Celsius, and cm). These are finite-run summaries of the station-maximum diagnostic.}
\label{tab:maximum-uplift}\label{tab:station-set-accuracy}
\begin{tabular}{llrrrrr}
\toprule
Dataset & Method & $K$ & $S$ & $U$ & \shortstack{Largest\\station mean}
& \shortstack{Maximization\\contribution}\\
\midrule
UK & L-MQ & 0.080 & 0.947 & 0.045 & 180.658 & 5.042\\
& L-MQ-TMC & 0.012 & 0.965 & 0.032 & 182.094 & 5.526\\
& MCMC & --- & 1.000 & --- & 182.181 & 5.633\\
France & L-MQ & 0.294 & 0.847 & 0.102 & 43.159 & 1.077\\
& L-MQ-TMC & 0.122 & 0.871 & 0.057 & 43.288 & 1.129\\
& MCMC & --- & 1.000 & --- & 43.290 & 1.265\\
Canada & L-MQ & 0.053 & 0.940 & 0.040 & 664.360 & 1.858\\
& L-MQ-TMC & 0.010 & 0.964 & 0.031 & 677.355 & 1.926\\
& MCMC & --- & 1.000 & --- & 679.419 & 2.001\\
\bottomrule
\end{tabular}
\end{table}

\subsection{Monte Carlo variability of station maxima}
\label{app:precision}
Table~\ref{tab:paired-regions} reports paired Monte Carlo intervals for the maximum diagnostics using the resampling design of Section~\ref{sec:accuracy}.

\begin{table}[htbp]
\centering\small
\caption{Station-maximum circular-block bootstrap intervals using the design of Section~\ref{sec:accuracy}. Negative $\Delta K=K_{\mathrm{TMC}}-K_{\mathrm{MQ}}$ favors the correction. The final column is the central 95\% finite-pool noise range from the common-law benchmark. Intervals describe Monte Carlo variability conditional on the saved references.}
\label{tab:paired-regions}\label{tab:noise-benchmarks}
\begin{tabular}{llrrr}
\toprule
Dataset & TMC $K$ interval & $\Delta K$ & $\Delta K$ interval & Benchmark\\
\midrule
UK & $[0.009,0.020]$ & $-0.067$ & $[-0.075,-0.057]$ & $[0.005,0.013]$\\
France & $[0.115,0.132]$ & $-0.172$ & $[-0.177,-0.168]$ & $[0.005,0.017]$\\
Canada & $[0.007,0.016]$ & $-0.044$ & $[-0.049,-0.031]$ & $[0.004,0.014]$\\
\bottomrule
\end{tabular}
\end{table}

For the northern French station maximum, the L-MQ-TMC CDF discrepancy exceeds its finite-pool Monte Carlo benchmark from Section~\ref{sec:accuracy}. The corresponding UK and Canadian CDF discrepancies lie within their benchmark ranges (Table~\ref{tab:noise-benchmarks}).

\subsection{Support of the Gaussian reconstructions}
The original GEV likelihood has parameter-dependent support, whereas an unconstrained Gaussian reconstruction need not respect every observed-data support inequality. Table~\ref{tab:draw-support} checks all station--draw pairs using each station's observed minimum and maximum block values. Approximate joint draws are retained rather than silently rejected, because rejection would define a different approximation. The modal derivatives and correction use the exact supported likelihood; that fact does not guarantee supported Gaussian draws or calibrated approximate tails.

\begin{table}[htbp]
\centering\small
\caption{Approximate draws violating the observed-data GEV support at one or more stations, out of 20,000 draws. Pairs count station--draw combinations.}
\label{tab:draw-support}
\begin{tabular}{llrr}
\toprule
Dataset & Method & Joint draws & Station--draw pairs\\
\midrule
UK & L-MQ & 0 & 0\\
UK & L-MQ-TMC & 0 & 0\\
France & L-MQ & 837 & 859\\
France & L-MQ-TMC & 291 & 295\\
Canada & L-MQ & 0 & 0\\
Canada & L-MQ-TMC & 0 & 0\\
\bottomrule
\end{tabular}
\end{table}
\FloatBarrier
\end{document}